\documentclass{article}

\usepackage[english]{babel}

\usepackage[letterpaper,top=2cm,bottom=2cm,left=3cm,right=3cm,marginparwidth=1.75cm]{geometry}

\usepackage{amsmath}
\usepackage{graphicx}
\usepackage{adjustbox}
\usepackage[auth-sc, affil-it]{authblk}
\usepackage[colorlinks=true, allcolors=blue]{hyperref}
\usepackage[normalem]{ulem}

\usepackage{parskip}
\providecommand{\keywords}[1]
{
  \small	
  \textbf{\textit{Keywords---}} #1
}

\providecommand{\listofabbreviation}[1]
{
  \small	
  \textbf{\textit{List of abbreviations---}} #1
}

\title{Data-driven multi-objective optimization for alloy recycling using factorization machines and quantum annealing}

\author[1,*]{Thomas Plehn}
\author[1]{Katrin Bugelnig}
\author[1]{Silvana Tumminello}
\author[1]{Daniel Barragan-Yani}
\author[1]{David Melching}

\affil[1]{Institute for Frontier Materials on Earth and in Space, German Aerospace Center (DLR), Linder Hoehe, Cologne, 51147, Germany}
\affil[*]{Corresponding author: Thomas.Plehn@dlr.de}

\begin{document}
\maketitle

\begin{abstract}
 
Quantum annealing has the potential to provide practical quantum advantage for complex optimization tasks. Here, we present a systematic assessment of an integrated factorization-machine and quantum-annealing workflow (FM+QA) for a technologically relevant application: multi-objective Pareto optimization in metal up-cycling through alloy design. To address the non-convex nature of the Pareto front, we employ the recently proposed data-driven Tchebycheff scalarization (DDTS) scheme. Our results show that FM+QA extends the applicability of QUBO-based optimization to data-driven materials discovery problems with multiple competing objectives. In particular, we analyze the scaling behavior of the approach and compare quantum annealing with classical simulated annealing using both regular binary encoding and one-hot encoding. Finally, we provide a critical perspective on the problem sizes and encoding strategies for which quantum-annealing-based optimization may become practically beneficial in the near future.

\end{abstract}


\listofabbreviation{QUBO: Quadratic unconstrained binary optimization, FM: Factorization machine, QA: Quantum annealing, SA: Simulated annealing, TC: Thermal conductivity, YS: Yield strength, TTS: time-to-solution, SOO: single-objective optimization, MOO: Multi-objective optimization}

\section{Introduction}
Globally, societies are facing socio-political challenges, such as supply bottlenecks, heavy reliance on critical resources and the need to achieve carbon neutrality~\cite{Scheffler2022-ot,Raabe2023-nf}. The aerospace and transport sectors require large quantities of metals for infrastructure and vehicles. However, producing primary materials, particularly aluminium and steel, is highly energy-intensive and results in high CO$_2$ emissions. These issues can be mitigated and sustainability improved by placing greater emphasis on recycled alloys. For instance, producing recycled aluminium requires only 5\% of the energy needed for primary aluminium production~\cite{IEA}. 
    
It is therefore necessary to analyse and develop alternative recycling routes. Secondary materials can be utilised by blending various types of scrap from different sectors and sources to create new alloys for specific manufacturing processes and applications. This can be realized by a combination of computer-based predictions and experimental investigations, thereby significantly reducing the development time for new alloys. For instance, J. Gao et al.~\cite{Gao2023-pn} developed an accelerated design strategy for hypereutectic Al–Si–Mg–Sc alloys by combining high-throughput CALPHAD simulations with Bayesian optimization. Microstructure–property relationships were established via active learning from experiments. More recently, Bugelnig et al.~\cite{Bugelnig2026preprint} demonstrated a scrap-based alloy discovery workflow for LPBF-processable Al--Si--Cu--Mg--Ni alloys, combining high-throughput thermodynamic screening, machine learning-assisted property prediction, optimization, uncertainty analysis, and synchrotron-based experimental validation. This experimentally validated workflow highlights the technological relevance of scrap-mixture alloy design, while the present study focuses on the complementary question of how data-driven QUBO-based active learning and quantum annealing can be used to explore such multi-objective recycling design spaces. Furthermore, Sánchez et al.~\cite{Sanchez2025-vm} developed a strategy to increase scrap recycling rates and produce low-carbon Al-castings by combining high-throughput CALPHAD-based simulations with experiments. They designed a multicomponent Al--Si--Cu--Zn--Fe--Mn--Mg alloy from post-consumer scrap by optimizing the solidification interval to promote eutectic formation and suppress primary intermetallic phases. The results were validated through microstructural and mechanical characterization.

On the computational side,  quantum technologies are being explored as accelerators for selected computational tasks in materials science. More specifically, quantum annealing (QA), a specialized approach to quantum computing, has been investigated as a promising heuristic for complex optimization problems~\cite{King2025, Kim2025, Munoz2025}. It does so by leveraging quantum mechanics, specifically quantum tunneling, adiabatic evolution and superposition effects, to sample low-energy configurations in large discrete search spaces~\cite{ Kadowaki1998,  Hauke_2020, Yarkoni2022}. From a practical point of view, the central idea of QA-based optimization is that entangled qubits naturally evolve into the ground state of a problem Hamiltonian represented by a corresponding quadratic unconstrained binary optimization (QUBO) problem. The latter is a rather restrictive formulation when trying to tackle general optimization tasks, but recent efforts have shown that machine learning-based approaches can help modelling realistic use-cases in this form. Particularly promising is the combination of QA with the idea of a factorization machine (FM)~\cite{Rendle2010}, in the so-called FM+QA (or FMQA) approach~\cite{tamura2025}. In the area of material science, the FM+QA method has already shown great potential as a single-objective black-box optimizer, e.g., applied on optical multilayer systems and metamaterials~\cite{Kitai2020, Kim2022, Kim2023}, core-shell nanoparticles~\cite{Urushihara2023}, the optimization of atomistic systems~\cite{Nawa2023}, for lattice prediction and optimization~\cite{ CouzinieJPSJ2025, XuNPJ2025}, and for the design of transparent radiative coolers~\cite{Kim2022, Hwang2025}.
Recently, a data-driven Tchebycheff scalarization (DDTS) scheme was introduced in \cite{Plehn2025} to extend data-driven QUBO-based active learning beyond single-objective optimization and enable the discovery of non-convex Pareto fronts. This capability is particularly relevant for multi-phase alloy design, where competing properties are governed by nonlinear composition–property relationships.

\begin{figure}[htbp]
    \centering
    \includegraphics[width=0.95\columnwidth]{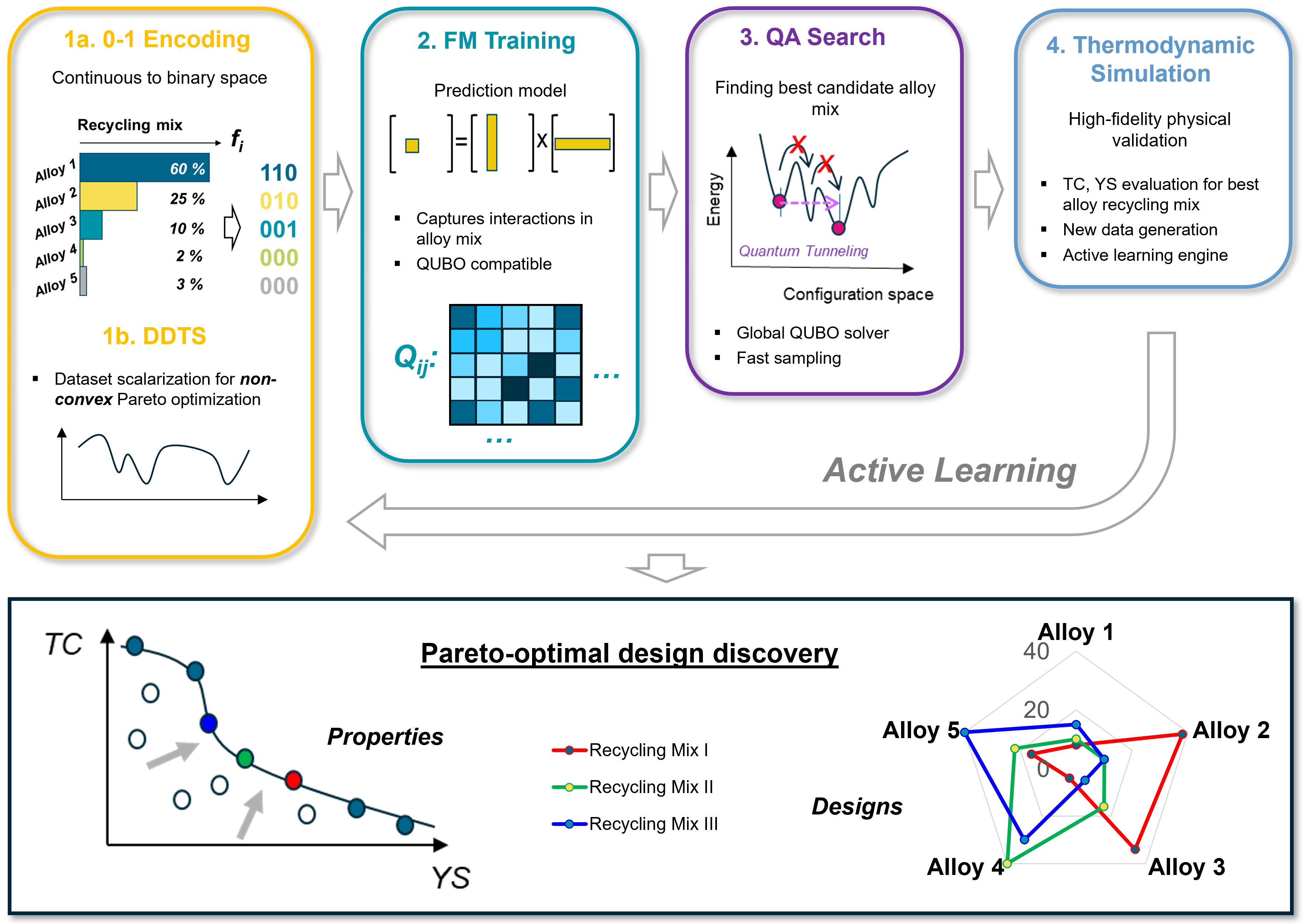}
    \caption{Factorization machine quantum-annealing (FM+QA) methodology with data-driven Tchebycheff scalarization (DDTS) for  non-convex multi-objective black-box optimization.  The active learning framework is applied to an alloy recycling use case in which new alloy mixtures are sought with Pareto-optimal combinations of target properties. Specifically,  thermal conductivity (TC) and yield strength (YS) are maximized by varying the blending ratios of five mixing alloys (\textit{Alloy 1}, \textit{Alloy 2}, \textit{Alloy 3}, \textit{Alloy 4}, \textit{Alloy 5}). Each active learning iteration starts with  dataset preprocessing, comprising 0-1 encoding of the alloy mixture ratios (step 1a) and  DDTS-based scalarization of the multi-objective property data (step 1b). Subsequently,  a factorization machine surrogate model is trained to predict the scalarized objective (step 2) and translated into a QUBO, or equivalently an Ising model. In step 3,  low-energy states of the corresponding Ising Hamiltonian are sampled by QA to propose a candidate alloy mixture. In step 4,  thermodynamic simulations provide the reference TC and YS values for the proposed mixture. The resulting alloy design is then  added to  the dataset, closing the active learning loop.  By varying the Pareto preference across iterations, the workflow enables the discovery of non-convex Pareto fronts.}
    \label{fig_figure1}
\end{figure}

This work implements and evaluates DDTS-enabled FM+QA for non-convex multi-objective Pareto optimization on quantum annealing hardware, using alloy recycling as a technologically relevant test case. As illustrated in Fig.~\ref{fig_figure1}, the workflow  extends data-driven QUBO-based active learning from single-objective optimization (SOO) to multi-objective Pareto optimization (MOO) by combining factorization machine surrogate models with the recently developed DDTS technique and solving the learned QUBO model on a quantum annealer \cite{Plehn2025}.  Such non-convex MOO problems are central to alloy design for aviation, where competing properties often arise from nonlinear composition--property relationships. Rather than treating QA as a standalone optimizer, we  assess its practical performance and limitations as a sampler within the active learning loop. Specifically, we compare binary and one-hot encoding strategies and evaluate FM+QA across small to large problem instances on the D-Wave Advantage hardware system \cite{Munoz2025, dwave_ocean_docs, dwave_advantage}, using the related classical FM+SA workflow based on simulated annealing (SA) as a reference. We apply the method to a recycling use-case of technological relevance: the optimization of Al-alloys suitable for additive manufacturing derived from mixtures of currently available Al-scrap compositions. Alloy mixtures comprising up to five alloys in varying ratios are considered. As an approximation, all alloys are represented by their nominal compositions, i.e., without accounting for impurities arising from prior service history or recycling streams.

The presented active learning framework starts from a small dataset of randomly generated alloy mixtures and iteratively proposes new next candidate mixtures. Ultimately, iterative experimental high-throughput validation of the latter would be desirable, however, all alloy properties in this study are determined from computational models for practical reasons.
Accordingly, the term \textit{material properties} refers to model-based estimates derived within the applied computational validation scheme, rather than experimentally measured quantities. In this regard, the present work focuses on a purely computational high-throughput framework for rapidly screening the modeled design space, identifying key property trends and selecting a reduced set of promising scrap mixtures. A final experimental verification is essential but lies beyond the scope of this study focused on methodological demonstration. 

The computations comprise a multitude of thermophysical, mechanical and thermodynamic alloy properties and are computed using \textit{Thermo-Calc} software \cite{Andersson2002-uu, thermocalcThermodynamicDatabases}. The calculation settings are configured to approximate additive manufacturing conditions, characterized by rapid solidification and reduced grain sizes compared to conventional as-cast states.
In particular, yield strength (YS) and thermal conductivity (TC) were selected as optimization objectives. Both are typically conflicting, i.e., a set of Pareto-optimal solutions is expected.
It should be noted, that predicting mechanical strength of complex alloy systems such as those investigated here remains inherently challenging, and the model-based estimates may deviate from experimental measurements. In particular, the physical models employed here account not for all factors affecting alloy strength. While they capture the strength contribution of the precipitation-hardenable Al-FCC matrix, they do not consider parameters such as the morphology and interconnectivity of the primary phases. These microstructural features have been shown to have an important effect \cite{Bugelnig2018, Bugelnig2023-wr}.
Although additional material properties could be included, in this work the objective space is deliberately kept simple in order to assess the performance of the optimization framework through three sources of complexity: the number of mixing alloys, the resolution of the alloy mixtures, and the employed 0-1 encoding scheme.

By demonstrating FM+QA applied to MOO problems, our study goes beyond recent state-of-the-art publications on the FM+QA method for material science.  While the DDTS scheme for multi-objective optimization with FM+QA was developed in our previous work \cite{Plehn2025}, its implementation on quantum annealing hardware has not yet been demonstrated. In the work of Kitai et al.~\cite{Kitai2020}, FM+QA was used to optimize a tiled multilayer metamaterial with respect to a single nonlinear figure of merit. The authors showed that, for small system sizes, FM-based modelling performed slightly better than a Gaussian process approach. For larger systems, however, the performance of FM+QA was only compared with a random search baseline. In Nawa et al.~\cite{Nawa2023}, FM+QA was applied to optimize an atomistic magnetic tunnel junction. Their study discussed several factors influencing the FM+QA method, but not the impact of the chosen 0-1 encoding, as the atomistic system model naturally prescribed the encoding. Although the authors compared FM+QA with FM+SA, i.e., simulated annealing (SA), and Bayesian optimization, the comparison did not extend to large-scale problems, since their model involved only 10 qubits.
Overall, our research contributes to a deeper understanding of the FM+QA approach and its potential applications in digital materials design and discovery.

\section{Results}
\label{sec_res}

We assess alternative recycling scenarios in order to define currently available Al-based scraps that are suitable for mixing. Put simply, existing alloys are recycled and combined to design new alloys. The design of the new alloy is determined by fractions [\%] of the involved scrap metal types, which we call \textit{mixing alloys} further on.

For our studies, five different mixing alloys are selected resembling technically relevant Al-based alloys in the middle range of strength and thermal conductivity: 
\begin{enumerate}
    \item \textbf{\textit{Alloy 1}} (AlSi12Cu3Ni2Mg1Fe0.5) \cite{Bugelnig2023-wr}, 
    \item \textbf{\textit{Alloy 2}}  (AlSi1Mg1Mn1) \cite{belov2005chapter2},
    \item \textbf{\textit{Alloy 3}} (AlSi1Cu6Mg1Mn1) ~\cite{belov2005chapter5},
   \item  \textbf{\textit{Alloy 4}} (AlMn1) ~\cite{belov2005multicomponent} and
    \item \textbf{\textit{Alloy 5}} (AlSi13Cu1Ni1Mg1Fe1) ~\cite{Bugelnig2018,Miladinovic2022}
\end{enumerate}

Moreover,  the alloy discovery process is guided by two target objectives: maximization of yield strength (YS) and maximization of thermal conductivity (TC). This renders the problem a  multi-objective Pareto optimization task in which no single alloy mixture is expected to optimize all objectives simultaneously.  Notably,  both target properties are expected to depend nonlinearly on the mixing alloy fractions, which makes the optimization task inherently complex.  The complexity of the resulting search space is controlled by two distinct factors: the resolution of the mixing alloy fractions [\%] and the number of mixing alloys included in the recycling system.  In this work, we examine both influences separately. Specifically, our analysis starts by comparing the FM+QA active learning framework with its classical counterpart, FM+SA, and by investigating the influence of the employed 0-1 encoding strategy. For this purpose, a SOO study aiming for YS maximization of the alloy mixtures serves  as a controlled benchmark with a known optimum.  Across the different sub-studies,  we systematically vary the resolution of the fractions and the number of mixing alloys to assess the method across multiple problem scales.  Afterwards, we turn to a MOO scenario to identify the Pareto-optimal, i.e., non-dominated, mixture designs that maximize both YS and TC.

\subsection{Single-Objective Optimization: Various Scales and 0-1 Encodings}
\label{sec_res_so}

Here, we  consider  SOO scenarios and aim for YS maximization driven through a maximum number of 150 active learning iterations, starting from a small initial dataset containing 10-20 random alloy mixtures. 
The performance of FM+QA is compared with FM+SA in five sub studies each characterized by different alloy recycling systems.

We define an alloy recycling system by the involved mixing alloys and their mixture resolution [\%].
The latter is necessary since  FM+QA works in  discrete search spaces, which here requires a discretization of the continuous variable mixing alloy fractions, as described in detail in Sec.~\ref{sec_methods_enc}.
According to the number of mixing alloys considered for recycling, we define sub studies and the corresponding recycling systems as: S (small, with 3 mixing alloys: \textit{Alloy 1}, \textit{Alloy 2}, \textit{Alloy 3}), M (medium, with 4 mixing alloys: \textit{Alloy 1}, \textit{Alloy 2}, \textit{Alloy 3}, \textit{Alloy 4}) and L (large, with 5 mixing alloys: \textit{Alloy 1}, \textit{Alloy 2}, \textit{Alloy 3}, \textit{Alloy 4}, \textit{Alloy 5}). 
Varying additionally the resolution of the fractions provides us with each two S and M sub studies, while only a single L system is investigated (see Table~\ref{tab_1}).

The consumption of logical qubits of each recycling system equals the number of required 0-1 variables for encoding. Therefore, it depends on the number of alloys, fraction resolution and crucially the employed type of 0-1 encoding. Table~\ref{tab_1} summarizes numbers of qubits required within each sub study. In particular, binary encoding exhibits the expected logarithmic scaling with increasing resolution, whereas one-hot encoding scales linearly and therefore requires substantially more qubits. Owing to the limited number of qubits available on current hardware, the resolution of the fractions was chosen comparatively coarse in the present study to keep the resulting QUBO instances computationally tractable.

\begin{table}[h!]
\centering
\begin{adjustbox}{width=1\textwidth}
\begin{tabular}{ |c||c|c|c|c|  }
 \hline
 \multicolumn{5}{|c|}{Alloy recycling models of different scales} \\
 \hline
 sub study &  Mixing alloys [\#] & Resolution [\%] & Binary encoding [\# qubits] & One-hot encoding [\# qubits] \\
 \hline
 S1 & 3 & 14.3 & 9  & 21 \\
 S2  & 3 & 6.6 & 12 & 45 \\
 M1 & 4 & 14.3 & 12 & 28 \\
 M2 & 4 & 6.6  & 16 & 60 \\
 L  & 5& 3.1  & 25 & 155* \\
 \hline
\end{tabular}
\end{adjustbox}

\caption{Details on five different alloy recycling systems introduced for SOO (maximization of YS). Each line summarizes a sub study indicated from S1 (3 mixing alloys) to L (5 mixing alloys). Resolution of alloy mixtures defines the required number of qubits for 0-1 encoding, specifically for binary and one-hot encoding. \\
*) The L mixture system with one-hot encoding is filed for completeness, although the simulations were out of scope.}

\label{tab_1}
\end{table}

By comparing FM+QA and FM+SA with both binary and one-hot encoding, we assess the impact of the sampler and the encoding strategy within an automated high-throughput active learning workflow. For consistency, throughout all of our studies, the number of runs and sweeps used for classical SA within the FM+SA algorithm are kept unchanged, while rigorous adjustments of the QA device are considered, as described in Sec.~\ref{sec_methods_fmqa}.

To assess the variability of the active learning performance, all optimization runs were repeated 15 times with independently sampled random initial datasets of identical size. We report the resulting mean values and standard deviations.

YS maximization is the only objective of our recycling cases in this part. According to our computational models,  based on \textit{Thermo-Calc}, \textit{Alloy 1} provides highest YS among all of our pure alloys. Consequently, we expect the optimization process to return always the mixture ratio of 100 \% \textit{Alloy 1}.  Importantly, the pure mixing alloys are excluded from all initial datasets, so the optimization method has no prior knowledge of the globally optimal pure-alloy designs. Table~\ref{tab_2} in Sec.~\ref{sec_methods_data} provides the YS values for all five mixing alloys for comparison.

\subsubsection{Three alloy mixtures}

\begin{figure}[htbp]
    \centering
    \includegraphics[width=0.49\columnwidth]{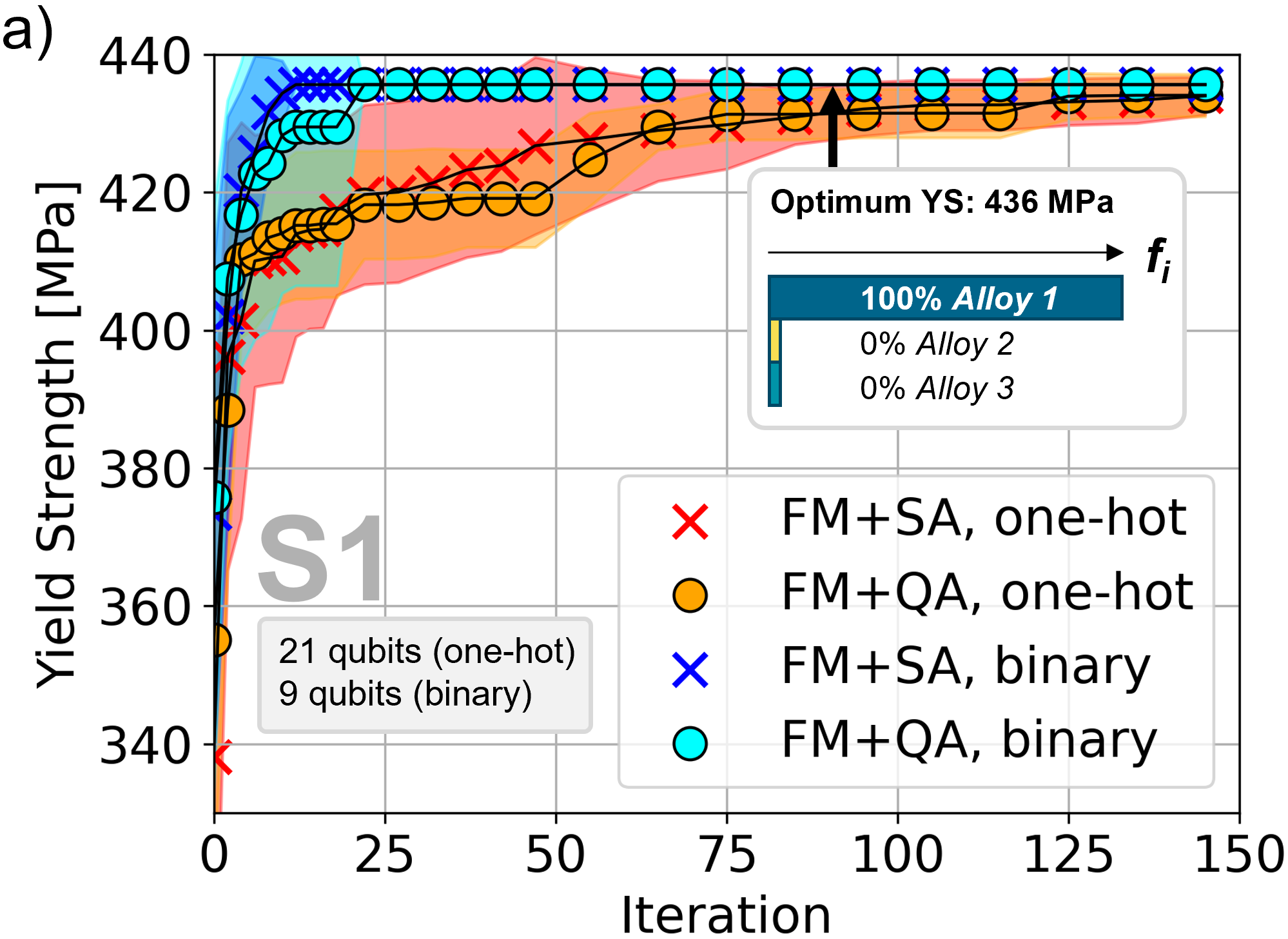}
    \includegraphics[width=0.49\columnwidth]{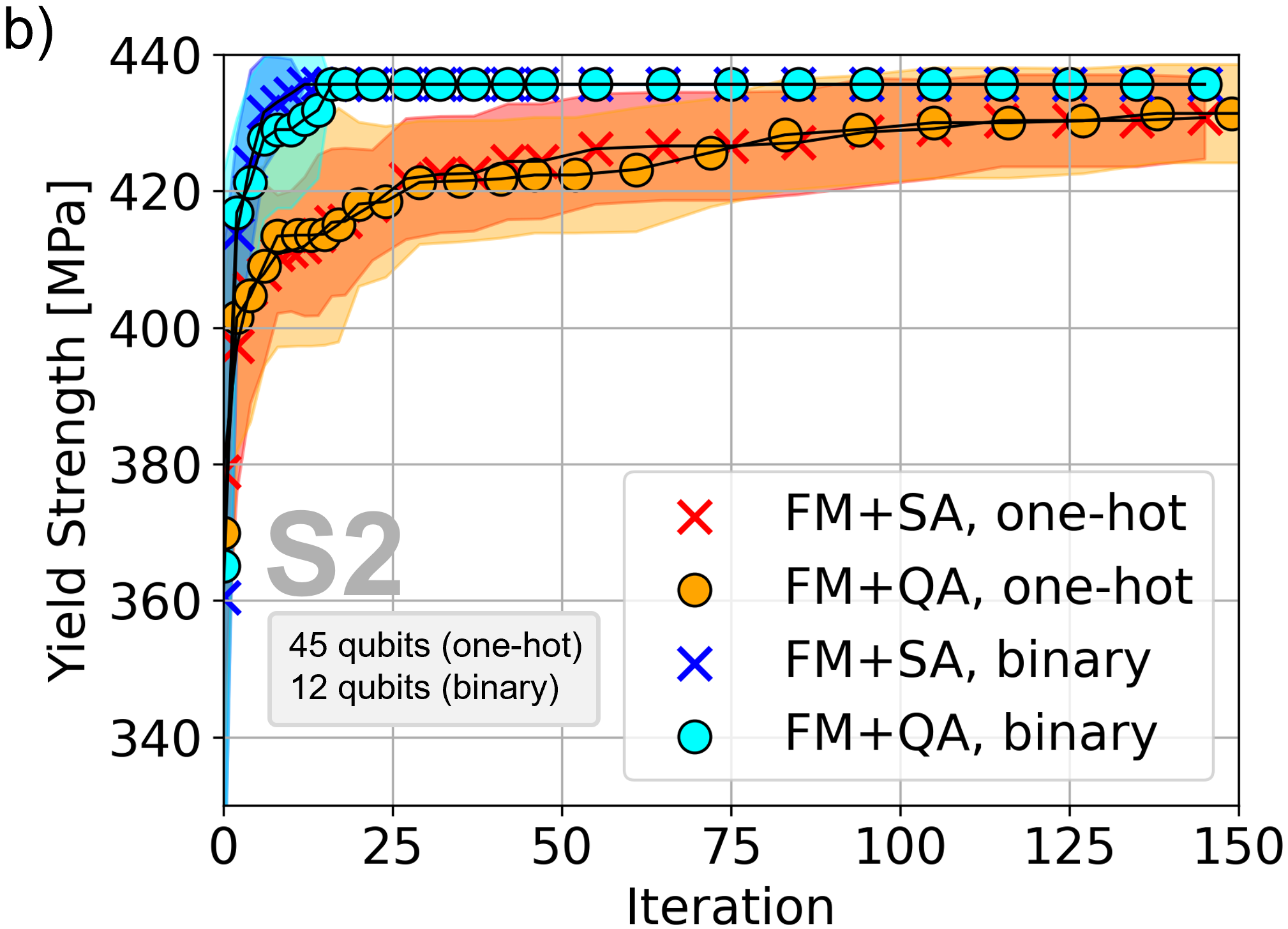}
    \caption{Yield strength maximization in the 3 mixing alloy models S1 (left) and S2 (right). Accumulated best mixtures found within 150 active learning iterations. 
    FM+QA is compared with FM+SA employing binary and one-hot encoding. Model S1 requires 9 (binary) or 21 (one-hot) qubits. S2 is based on 12 (binary) or 45 (one-hot) qubits. Marks and shadings represent mean values and standard deviations, respectively.
     }
    \label{fig_results_so_1}
\end{figure}

Results from sub studies S1 and S2 are shown in Fig.~\ref{fig_results_so_1}a and \ref{fig_results_so_1}b, respectively. 
S1 and S2 are the two small-scale alloy recycling systems with 3 mixing alloys with 14.3 \% and 6.6 \% resolution of the mixing alloy fractions (see Table~\ref{tab_1}). Both contain the mixing alloys \textit{Alloy 1}, \textit{Alloy 2} and \textit{Alloy 3}. The 0-1 encoding the mixing alloy fractions consumes different numbers of qubits; for the S1 model it is $3 \times 3=9$ ($3 \times 7=21$) qubits, while S2 requires $3 \times 4=12$ ($3 \times 15=45$) qubits, based on binary (one-hot) encoding, respectively. 

Within each FM+QA iteration of S1, a number of 20,000 QA runs turned out to be sufficient to obtain optimization progress similar to the analog FM+SA simulations (see Fig.~\ref{fig_results_so_1}a). The same setting was also employed in S2 with binary encoding to conform the FM+QA and FM+SA results (see Fig.~\ref{fig_results_so_1}b, warm colors). In contrast, in S2 with one-hot encoding (Fig.~\ref{fig_results_so_1}b, cold colors), 45,000 QA runs were required for FM+QA to achieve performance comparable to FM+SA. In addition, quantum error-mitigation measures were applied in the form of 45 spin-reversal transformations. This increased computational effort reflects the sensitivity of current QA hardware to the size of the embedded physical Ising system. On the D-Wave Advantage Pegasus graph, our logical QUBO models with clique sizes of 9, 12, 21, and 45 required embeddings onto subgraphs comprising 18, 36, 80, and 260 physical qubits, respectively.

Furthermore, Fig.~\ref{fig_results_so_1} clearly identifies the binary encoding (cold colors) as the better choice compared with one-hot encoding (warm colors). The FM+QA and FM+SA methods with binary encoding locate the globally optimal alloy mixture (100 \% \textit{Alloy 1}) with the YS maximum value of 436 MPa in fewer than 25 iterations. In contrast, FM+QA and FM+SA based on one-hot encoding require substantially more iterations to converge. The likely explanation is as follows: during the initial period of 25 iterations, the FM regression model improves such that predictions within parts of the search space covered with data are learned properly. Beyond this stage, however, due to the very sparse one-hot encoding, the FM struggles to expand prediction power further to untapped search space areas. These areas appear notoriously with one-hot encoding in terms of many 0-1 variables being zero for all alloy instances in the dataset. Binary encoding, by contrast, provides a much more space-efficient data representation in a lower-dimensional search space and therefore does not suffer from this limitation to the same extent. Furthermore, the one-hot encoding is clearly more qubit intensive, which generally makes searching more difficult, independent on data availability. 
Contrarily, one might also expect that the denser information storage of binary encoding could challenge the limited second-order expressivity of the FM surrogate model. However, no such detrimental effect is observed in the present study.

\subsubsection{Four alloy mixtures}

 We next consider  the medium scale alloy mixing models, M1 and M2, based on four mixing alloys: \textit{Alloy 1}, \textit{Alloy 2}, \textit{Alloy 3} and \textit{Alloy 4} (see Table~\ref{tab_1}). Thereby, models M1 and M2 apply 14.3 \% and 6.6 \% resolution of the mixing alloy fractions, respectively, and consume $4 \times 3=12$ (M1, binary) or $4 \times 7=28$ (M1, one-hot) as well as $4 \times 4=16$ (M2, binary) or $4 \times 15=60$ (M2, one-hot) logical qubits. 

Furthermore, here we support the active learning framework when using one-hot encoding by introducing a local search mechanism based on spin-flip actions, applied whenever the QA or SA solver delivers a solution for a second time, i.e., an alloy design that is already in the dataset. The mechanism is specifically dedicated to mitigate the slow final convergence in all one-hot runs (FM+SA and FM+QA consistently) and acts as a local (exploitative) search in balance with a global (exploratory) search for random alloy mixtures. Please note, that this mechanism is not applied to the binary encoding runs as those do not suffer from premature convergence (see section above).

\begin{figure}[htbp]
    \centering
    \includegraphics[width=0.49\columnwidth]{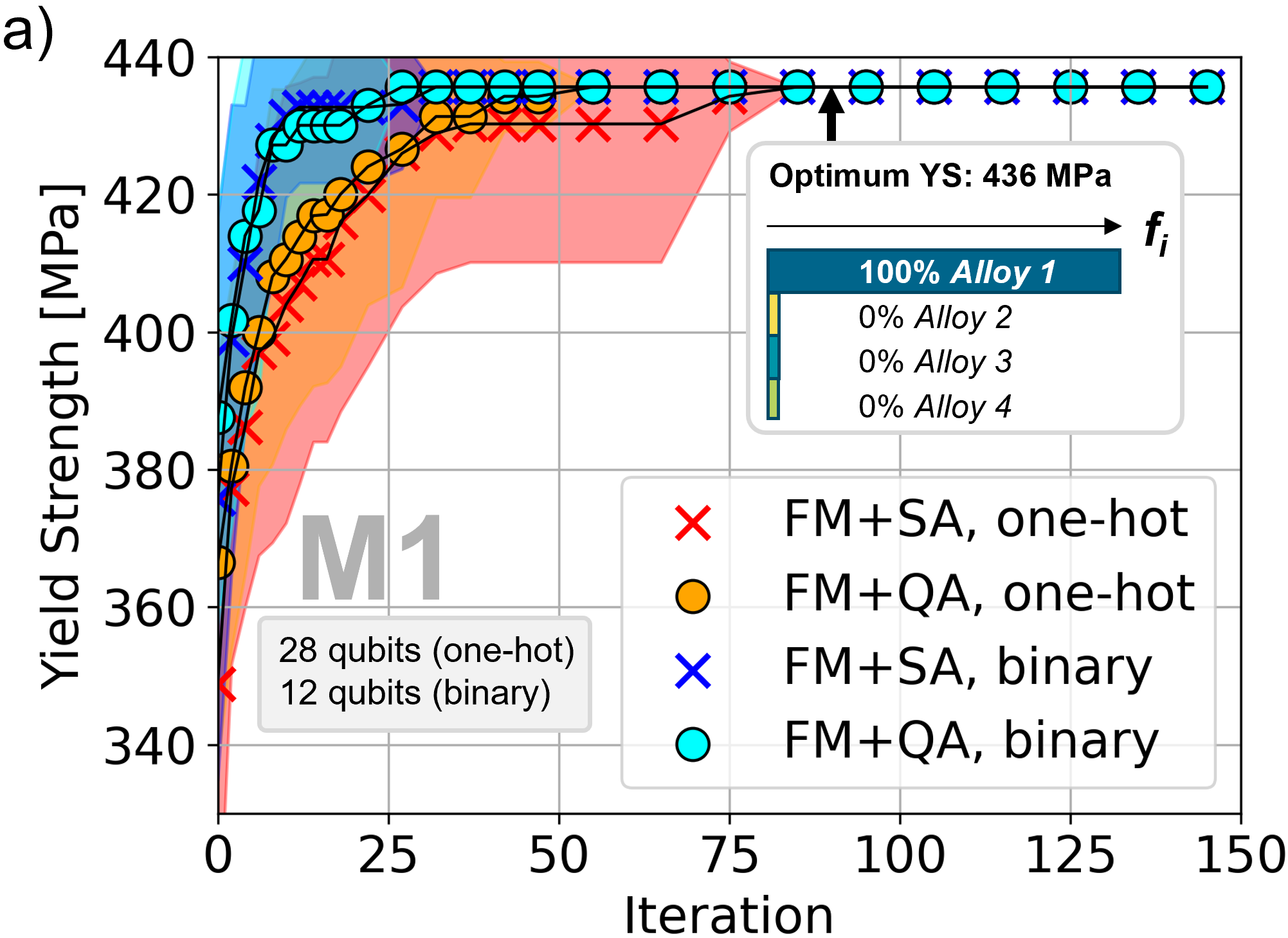}
    \includegraphics[width=0.49\columnwidth]{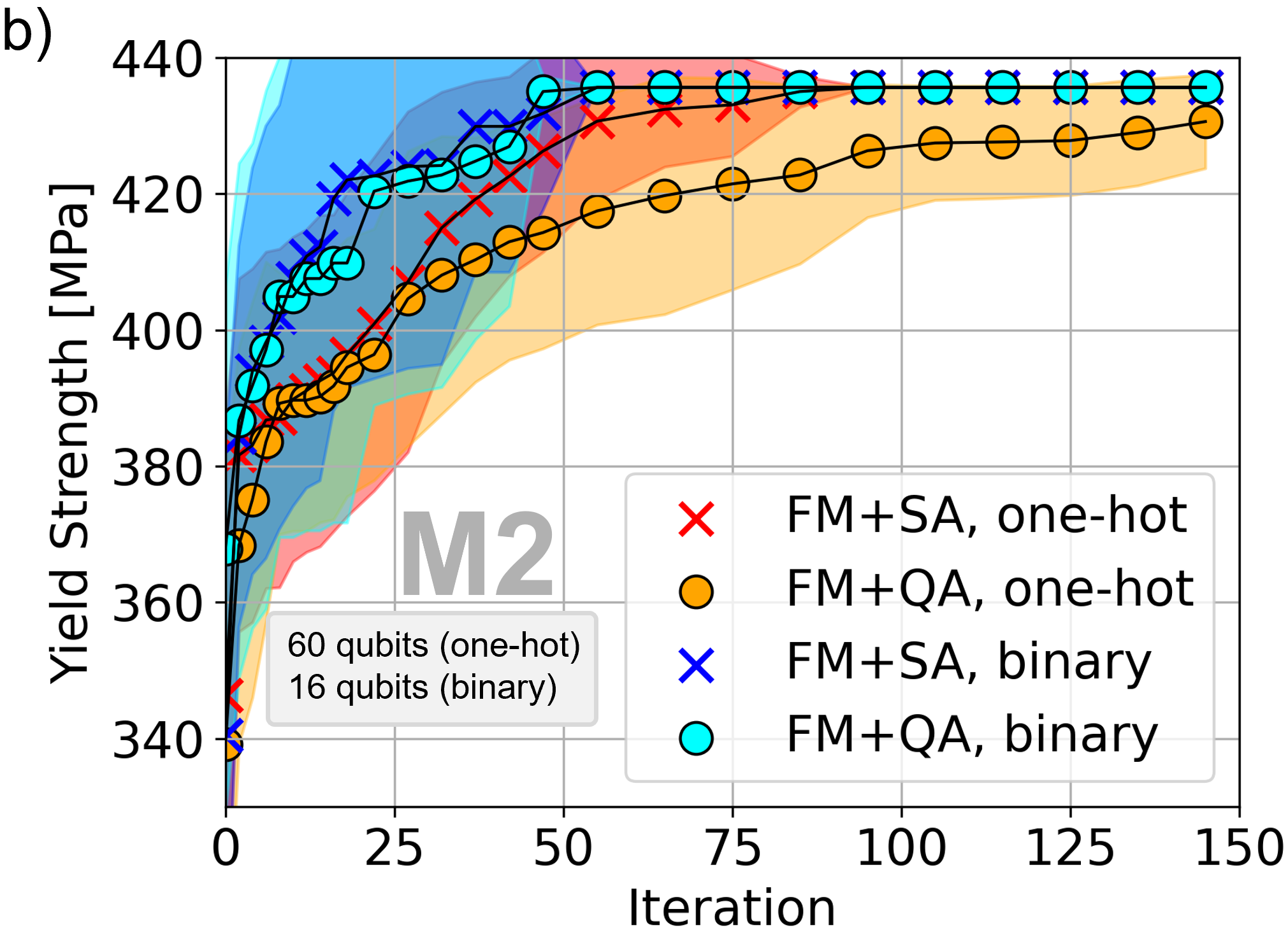}

    \caption{Yield strength maximization in the 4 mixing alloy models M1 (left) and M2 (right). Accumulated best mixtures found within 150 active iterations. 
    FM+QA is compared with FM+SA employing binary and one-hot encoding. Model M1 requires 12 (M1, binary) or 28 (M1, one-hot) qubits. M2 is based on 16 (M2, binary) or 60 (M2, one-hot) qubits. 
}
    \label{fig_results_so_2}
\end{figure}

Results from sub study M1 are illustrated in Fig.~\ref{fig_results_so_2}a. In each iteration of the FM+QA with one-hot encoding, the QA solver utilizes again 45,000 shots inclusive 45 spin-reversal transformations, while the FM+QA with binary encoding still runs on only 20,000 QA repetitions. The SA setting is kept unchanged with respect to S1 and S2. Again, we observe better performance of the active learning process based on binary encoding, where FM+QA and FM+SA demonstrate similar rapid convergence behaviour. FM+SA and FM+QA with one-hot encoding exhibit faster final convergence to the global optimal alloy design than observed before in S1 and S2 (see Fig.~\ref{fig_results_so_1}, warm colors), owing to the effect of the introduced local search engine. Further, in Fig.~\ref{fig_results_so_2}a we realize no significant difference between FM+QA and FM+SA based on one-hot encoding. Since the M1 model with one-hot encoding needs 28 qubits, this finding agrees well with our previous results of sub studies S1 and S2 (21 and 45 qubits, respectively).

Results from sub study M2 are provided in Fig.~\ref{fig_results_so_2}b. For both encoding implementations, the QA solver and active learning setup is now kept unchanged with respect to M1, although the number of 0-1 variables increases to 16 and 60 for binary and one-hot encoding, respectively. By that, it is our purpose to finally witness the deterioration of the FM+QA algorithm with one-hot encoding (Fig.~\ref{fig_results_so_2}b, orange circles) against its classical FM+SA analogue (Fig.~\ref{fig_results_so_2}b, red crosses). 
As the QA is running on NISQ hardware, the illustrated performance breakdown of FM+QA is expected for large qubit numbers. Here, we can identify this point between 45 and 60 qubits. Notably, embedding 60 fully connected logical qubits here stretches to an interacting spin system of 408 physical qubits within the QA device.
Regarding simulations based on binary encoding (cold colors), FM+QA and FM+SA deliver again similar results, not surprising according to the still small number of 16 qubits needed for encoding model M2.
Further, the active learning performance loss when turning from M1 to M2 as experienced for all runs is reasonable. The logical search space of model M2 contains 16x (binary) and 4,294,967,296x (one-hot) more possible spin states than the search space of M1, nevertheless, all M1 and M2 runs are conducted with analogue settings and initialize from the same datasets of only 20 random alloy mixtures.

\subsubsection{Five alloy mixtures}
\label{sec_res_so_mix_5}

Finally,  we consider  sub study L finding the optimal recycling ratio of the 5 mixing alloys \textit{Alloy 1}, \textit{Alloy 2}, \textit{Alloy 3}, \textit{Alloy 4} and \textit{Alloy 5}. The model L assumes 3.2 \% resolution of the mixing alloy fractions. Binary encoding requires $5 \times 5=25$ variables, while the corresponding one-hot encoding would be based on $5 \times 31=155$ logical qubits. The latter is close to the maximum clique size that can be embedded on the D-Wave Advantage QPU. Therefore, the simulation runs of the FM+QA with one-hot encoding are beyond scope, and we concentrate on FM+QA as well as FM+SA with binary encoding.

\begin{figure}[htbp]
    \centering
    \includegraphics[width=0.49\columnwidth]{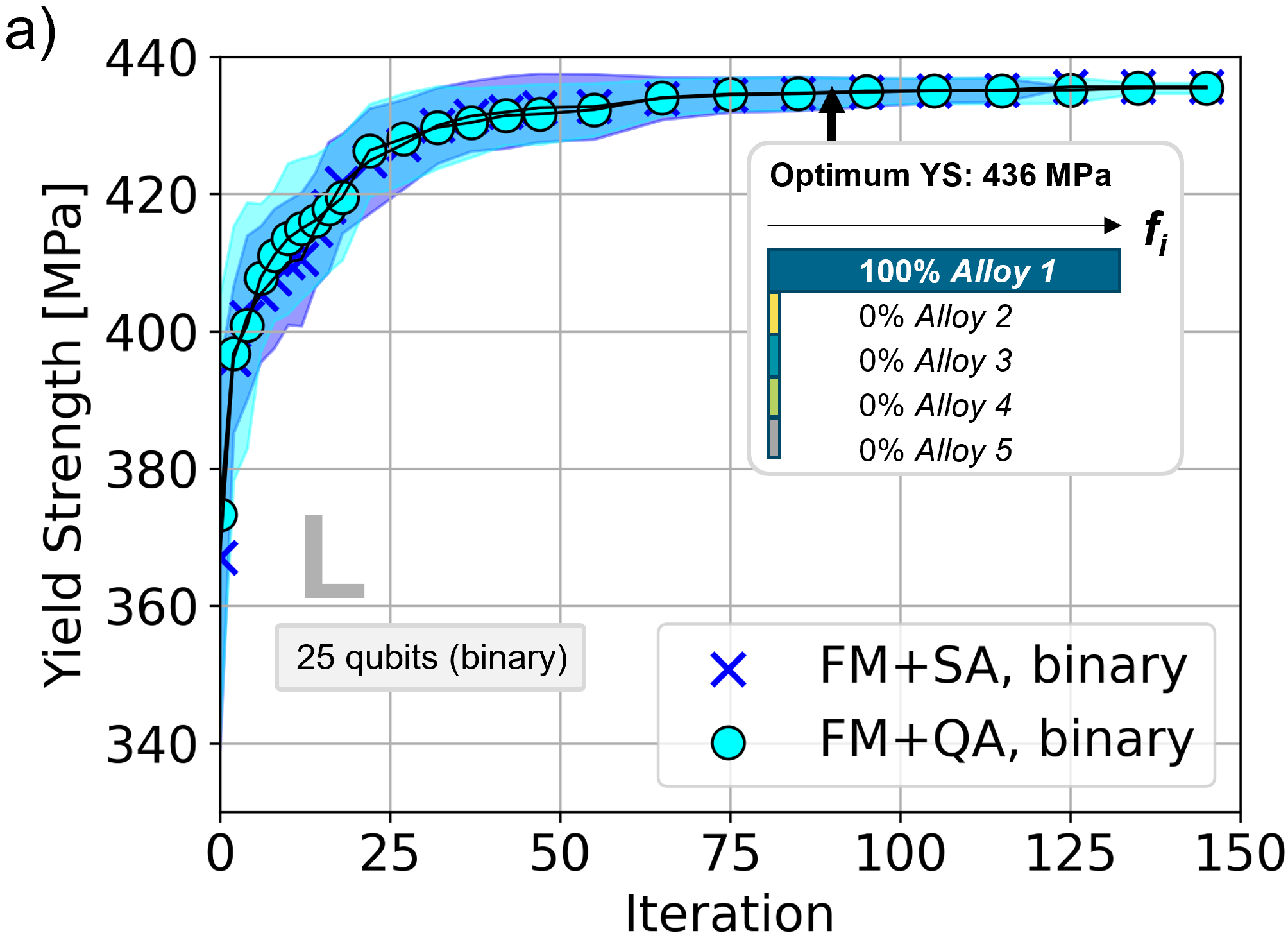}
    \includegraphics[width=0.49\columnwidth]{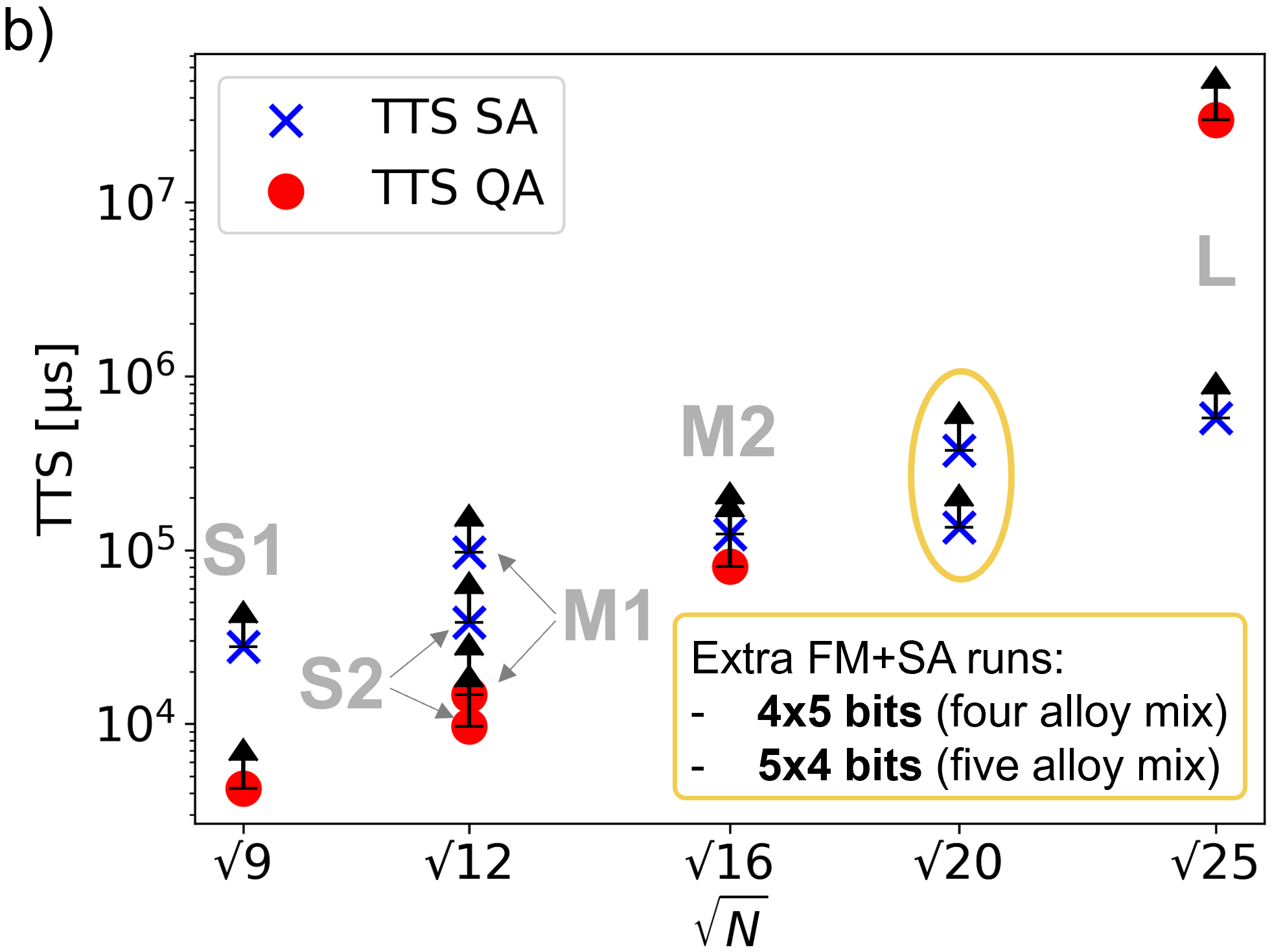}
    \caption{a) Yield strength maximization in the 5 mixing alloy model L1. Accumulated best mixtures found within 150 active iterations. FM+QA is compared with FM+SA based on binary encoding. Model L1 requires 25 qubits. b) Time-to-solution measure for QA and SA solvers as observed throughout our FM+QA and FM+SA with binary encoding active learning optimization framework. Different sub studies S1, S2, M1, M2 and L are indicated. Two more TTS values are plotted as observed from extra FM+SA runs for a quaternary alloy mixture model with 5 qubits resolution and a quinary model with 4 qubits.
   }
    \label{fig_results_so_3}
\end{figure}

Fig.~\ref{fig_results_so_3}a shows the results comparing the performance of the FM+QA with binary encoding with its classical peer. QA runs with 40,000 QA anneals and 40 spin-reversal transformations are added. Further, the advanced local search by spin-flipping is used to support the active learning (for FM+QA and FM+SA), however, the SA solver settings are kept again unchanged with respect to S1, S2, M1 and M2. The effect of error correction (in terms of the spin-reversal transformations) and the doubled number of annealing runs are the reason why we continue to observe a fairly similar performance of the active learning process compared to the corresponding M2 optimization, shown in Fig.~\ref{fig_results_so_2}b. Moreover, we observe a rapid initial optimization progress to somewhat slightly below 420 MPa. This behaviour can be explained by the structure of the objective landscape. In previous sub studies, the YS function sub-space for values above 338 MPa exhibits a ridge-like structure, where the primary ascent direction corresponds to increasing the fraction of  \textit{Alloy 1} in the mixture with the ridge maximum at 436 MPa. In the present case, however, YS up to 420 MPa can also be achieved by increasing the fraction of \textit{Alloy 5}. Consequently, in the range from 338 MPa to 420 MPa, the landscape transitions from a 1d ridge to an ascent 2d ridge surface spanned by increasing contributions of either \textit{Alloy 1} or \textit{Alloy 5} contributions. Optimization on this surface is less constraint and less fragile. Beyond this regime, starting from pure \textit{Alloy 5}, the landscape again reduces to a 1d ridge extending up to 436 MPa, and further improvement is again possible only along a single ascent direction (increasing \textit{Alloy 1}), while contributions of the remaining now 4 mixing alloys lead to decreasing YS.

We conclude that, with the present QA setup, FM+QA based on a 25 qubits binary encoding is again able to compare very well with corresponding FM+SA runs, overall in alignment with our previous sub studies.

\subsubsection{Time-to-Solution}

Time-to-solution (TTS), $\tau_{0.99}$, is the expected computational run time required to find the exact solution at least once with 99 \% probability, defined as
\begin{align}
\tau_{0.99} = \tau_{sr} \frac{\log (1-0.99)}{\log (1-p^*)},
\label{eq_tts}
\end{align}
taking into account computational time, $\tau_{sr}$, of a single run (i.e., cost for solving the QUBO once) and solution quality. Concerning the latter, the sampling process is considered as independent Bernoulli experiments with success probability, $p^*$, to find the exact QUBO solution. Here, we use $\tau_{0.99}$ to quantify the performance of the QA and SA solvers.

Please note that properly estimating the asymptotic scaling of the TTS for SA and QA methods crucially depends on applying the optimal annealing time 
\cite{Ronnow2014}. 
Therefore, thorough analysis of TTS scaling obligates problem-specific determination of the optimal annealing time for both solvers before recording solution runs for benchmarking, which is cumbersome within the automatized active learning context studied in this work.
Instead, here we provide TTS behaviour closer to practice, directly resulting from analysing $p^*$ of each FM+QA and FM+SA iteration (see Sec.~\ref{sec_methods_qa} for QA and SA solver adjustments).
For the SA solver, values for $\tau_{sr}$ refer to a single CPU of an Intel(R) Xeon(R) Gold 6242R processor with 3.10GHz. 
Concerning the QA solver, we conveniently utilize the QPU access time divided by the number of anneals for $\tau_{sr}$, for which estimates can be read out from the D-Wave QPU.

Fig.~\ref{fig_results_so_3}b visualizes $\tau_{0.99}$ of sub studies S1, S2, M1, M2 and L in dependence on model size in number of 0-1 variables, where the solvers face fully connected QUBOs of 9, 12, 12, 16 and 25 dimensions, respectively.
For practical reasons concerning the required exact QUBO solution references, we restrict the comparison to the optimization runs based on binary encoding.
 TTS measured on D-Wave Advantage hardware applied to solve 3D spin glass problems with randomized next neighbor couplings $J_{kl}\in\small\{-1,1\small\}$ and size $N_\text{qubits} = L \times L \times L$ for $L$ ranging up to 10 is reported to scale $\sim \exp (c \sqrt[3]{N_\text{bits}})$ 
\cite{King2020}, 
while SA is typically expected to provide TTS scaling $\sim \exp (c\sqrt{N_\text{bits}})$ on solving random fully connected QUBOs. While the latter trend of SA can be identified in Fig.~\ref{fig_results_so_3}b, the QA scaling measured here on random fully connected QUBO problem instances is worse than found in
\cite{King2020}, 
and notably also much less favourable than SA. Nevertheless, for our models S1, S2, M1 and M2 (i.e., up to 16 dimensions), the TTS of SA is an order of magnitude larger compared with QA.
The larger TTS in this area renders SA disadvantageous compared to QA in our specific high-throughput environment; however, due to lacking tuning of SA step number and sweep number, this does not apply to general statements.

\subsection{Multi-Objective Optimization}

\label{sec_res_mo}

 Next, we consider the MOO scenario where we aim to maximize YS as well as TC. The challenge in solving this Pareto optimization is that both properties usually conflict with each other, i.e., instead of a single ideal alloy mixture, we now search for the Pareto front as the complete set of non-dominated mixture designs \cite{Emmerich2018}. 

The alloy recycling system consists of the four mixing alloys already used in our previous sub studies M1 and M2: \textit{Alloy 1}, \textit{Alloy 2}, \textit{Alloy 3} and \textit{Alloy 4}. Here, however, we use a finer resolution of the fractions of 3.1 \%, which translates to a requirement of 20 qubits with binary encoding. We concentrate on binary encoding as one-hot encoding consistently results in worse performance in the above SOO scenarios (see Sec.~\ref{sec_res_so}). 
The QA computational costs per FM+QA iteration are 40,000 annealing runs and 40 spin-reversal transformations (similar to sub study L, discussed in Sec.~\ref{sec_res_so_mix_5}).  

\begin{figure}[htbp]
    \centering
    \includegraphics[width=\columnwidth]{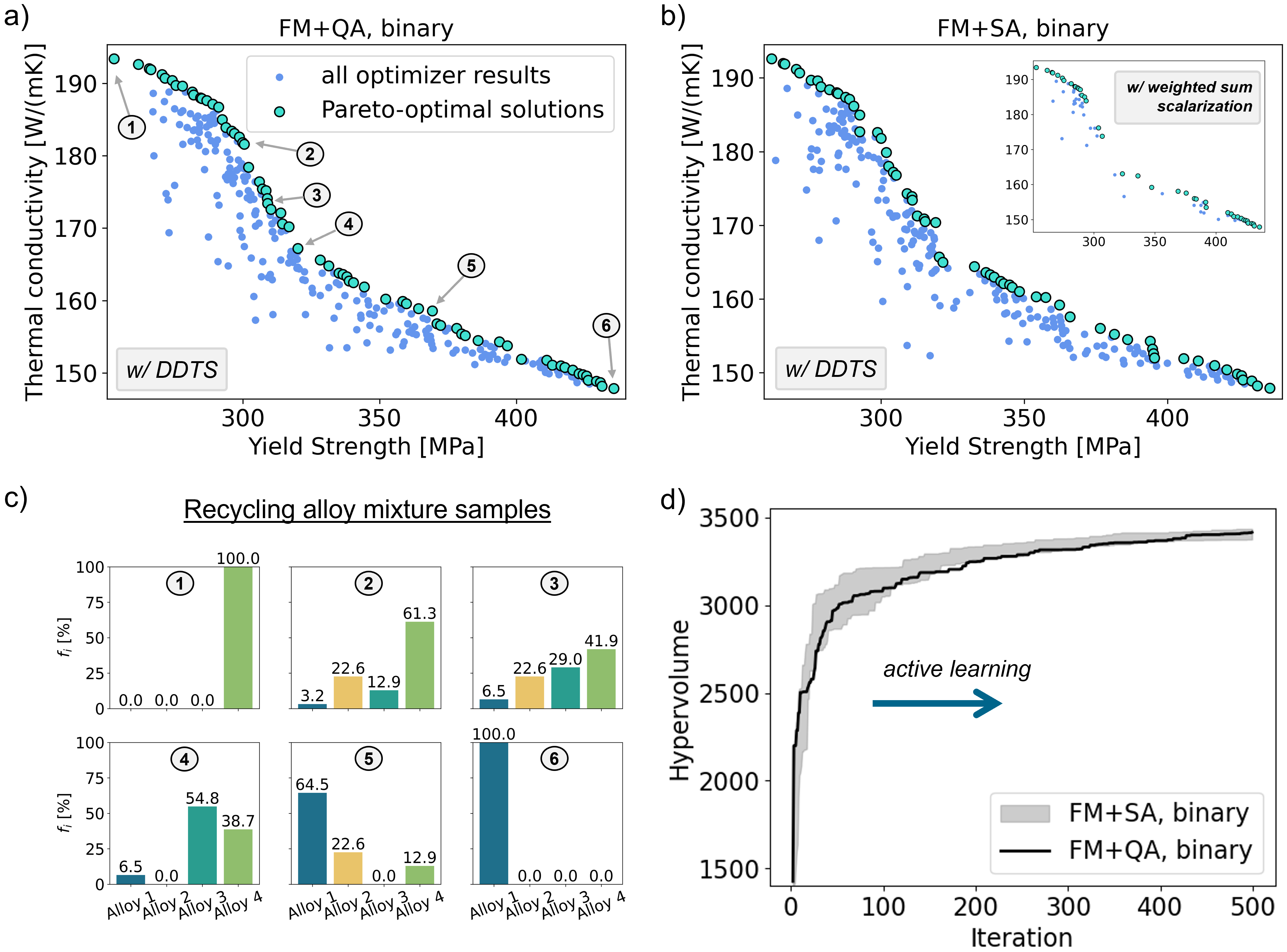}
    
    \caption{MOO results with the FM+QA (a) and FM+SA (b) algorithms. Objective space is spanned by the materials properties YS and TC, both to be maximized. The studied system consists of 4 mixing alloys with a resolution of 3.1 \%, which consumes in total 20 qubits. Each large cyan dot indicates a Pareto-optimal mixing ratio; the small light blue dots represent all solutions found by the QUBO optimization within 500 iterations. The inset in (b) gives corresponding FM+SA results utilizing weighted-sum scalarization instead of DDTS.
    c) Ratios of six selected Pareto-optimal recycling alloy mixtures from the FM+QA run in a). d) Iterative progress of the Pareto front hypervolume. The black line represents the active learning progress of the single performed FM+QA run in a). The gray shading, for comparison, shows the statistical variance of 5 classical FM+SA run repetitions.
   }
    \label{fig_results_mo_1}
\end{figure}

For Pareto front discovery, the FM+QA approach must be combined with QUBO scalarization techniques. For non-convex Pareto fronts or Pareto fronts with non-convex regions, however, applying a standard weighted-sum method to learned single-objective QUBO models is incapable of resolving the entire front \cite{Pardalos2017, Pereira2022, Ikeda2024}, critically restraining the effectiveness of the FM+QA approach. Here, we therefore introduce our data-driven Tchebycheff scalarization (DDTS) framework which implements the scalarization in terms of a dataset preprocessing step and works very well for uncovering also non-convex parts of a Pareto front \cite{Plehn2025}.

Fig.~\ref{fig_results_mo_1}a shows the Pareto front in 2D objective space spanned by our two objectives, YS and TC, as the result of a single FM+QA with DDTS run consisting of 500 iterations. The active learning initializes from 20 random alloy mixtures and iteratively searches for improving already found Pareto-optimal mixtures, leading to the construction of the shown Pareto front. 
As in previous single-objective studies, the pure mixing alloys are thereby not included in the initial dataset and thus unknown to the optimization method. The DDTS approach guarantees a Pareto front very well resolved along its entire course, notably also in the non-convex inner region. The homogeneous solution density along the entire front is obtained by shuffling the Pareto preference weights for each active learning iteration. This performance is exclusive for FM+QA with DDTS and cannot be achieved by applying the conventional weighted-sum scalarization method as demonstrated in the inset of Fig.~\ref{fig_results_mo_1}b. 
Clearly, the inner region of the Pareto front is only sparsely covered with Pareto-optimal solutions, as the weighted-sum scalarization fails in such non-convex regions \cite{Pardalos2017, Pereira2022, Ikeda2024}.

Fig.~\ref{fig_results_mo_1}b shows the result of an analogous run using the classical FM+SA approach. 
The comparison of Fig.~\ref{fig_results_mo_1}a and b demonstrates that, also when searching for Pareto-optimal solutions, no significant difference is observed between the FM+QA and FM+SA. To further substantiate this observation, Fig.~\ref{fig_results_mo_1}d shows the active learning progress in terms of the Pareto-front hypervolume for the FM+QA run (black line). An increasing hypervolume serves as a quantitative measure of progress in Pareto optimization. The gray shaded region in Fig.~\ref{fig_results_mo_1}d indicates the min–max range obtained from five independent FM+SA runs performed for comparison. Evidently, the optimization trajectory of the FM+QA approach lies well within the statistical variation of the classical runs.

\begin{figure}[t]
    \centering
    \includegraphics[width=0.49\columnwidth]{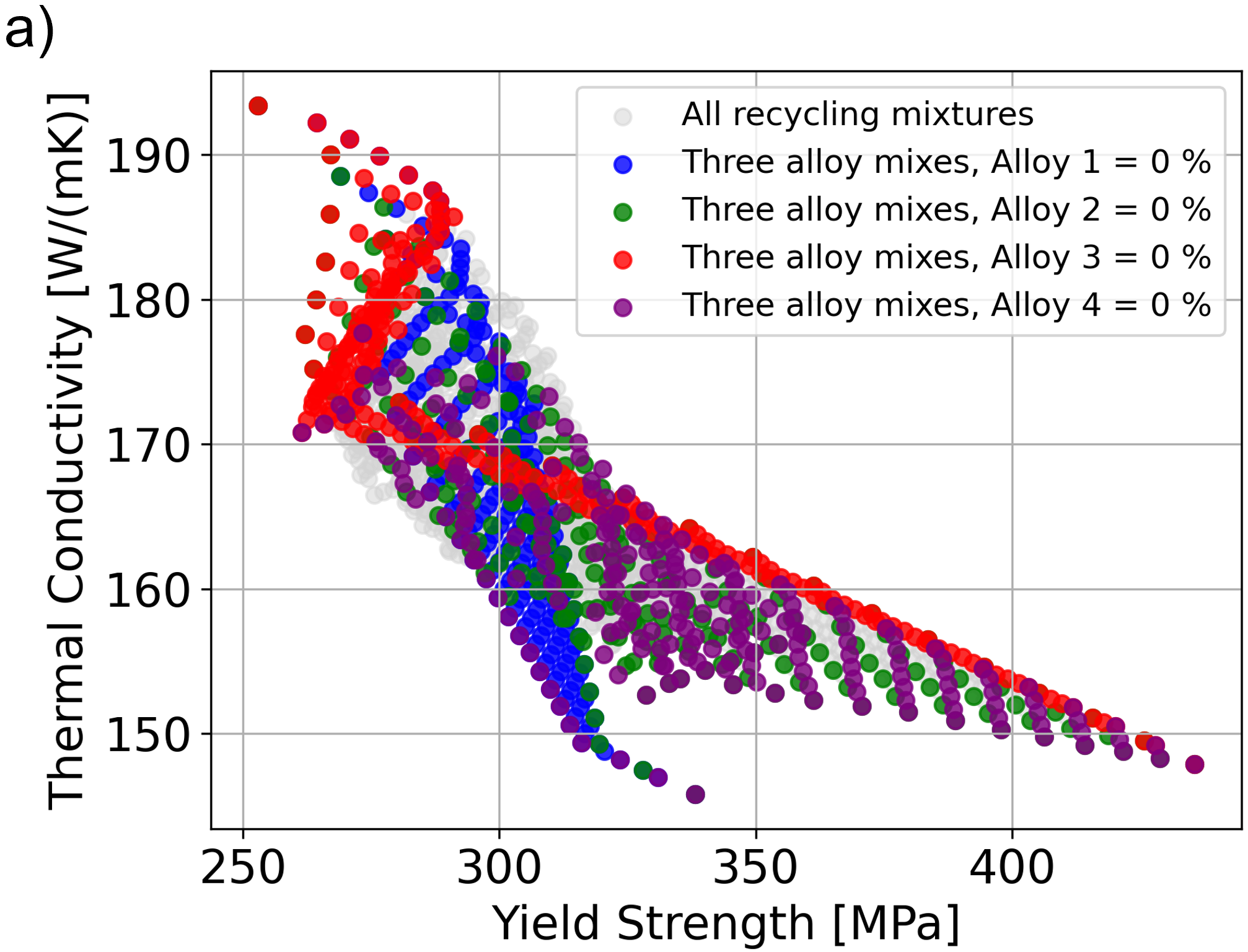}
    \includegraphics[width=0.49\columnwidth]{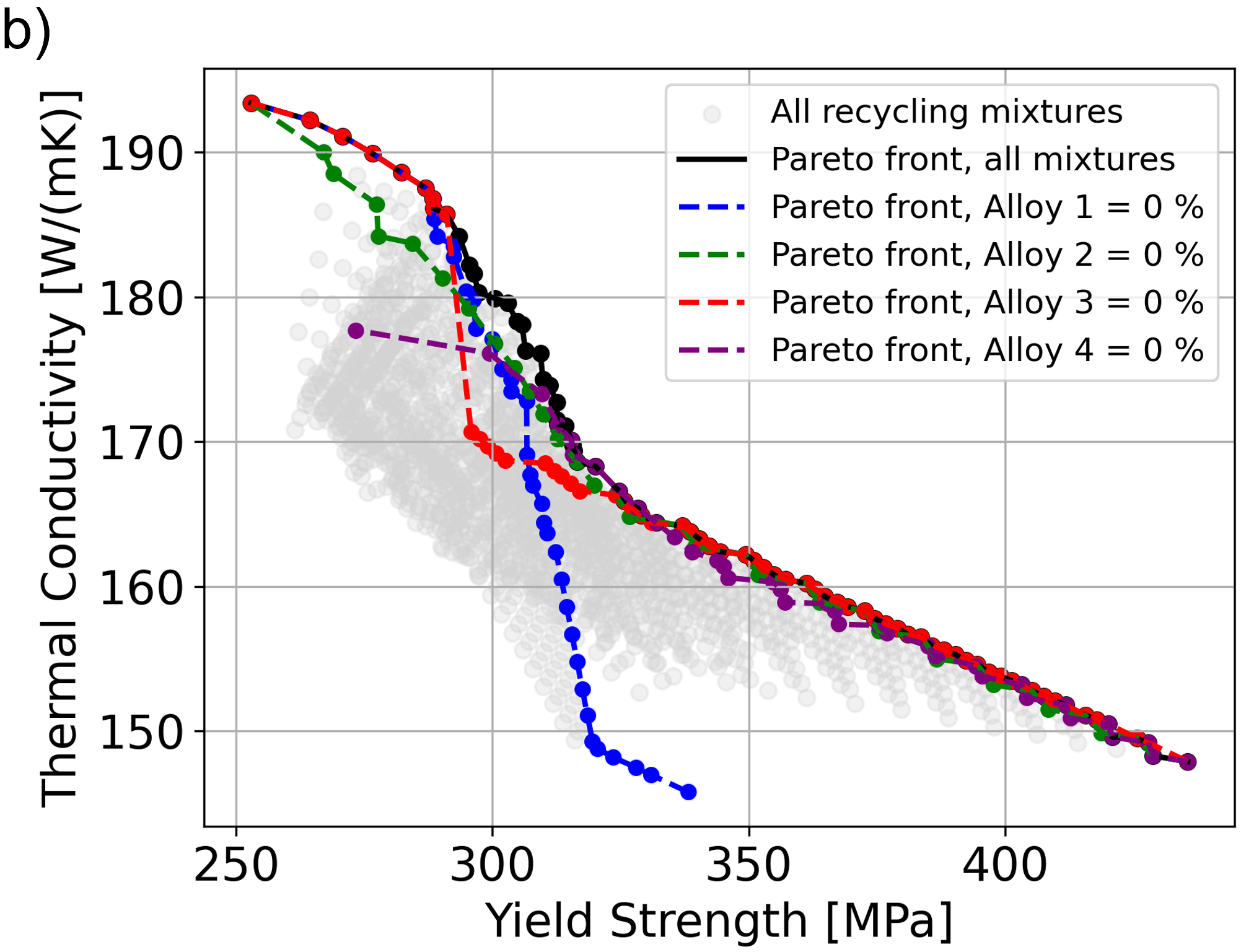}
    \caption{Alloy recycling mixtures obtained by brute force grid search applying a search space resolution of 5 \% for all mixing alloy fractions. a) All quaternary mixture designs in objective space with the four types of ternary mixtures highlighted. b) Same as a) visualizing the Pareto fronts of each data sub set.
}
    \label{fig_results_mo_3}
\end{figure}

 Finally, we  analyse the Pareto results in detail. As indicated in Fig.~\ref{fig_results_mo_1}c, the pure \textit{Alloy 3} and the pure \textit{Alloy 1} are returned as the upper and lower end point of the Pareto front, respectively, as expected according to Table~\ref{tab_2}. In between more versatile recycling alloy mixtures participate to the Pareto front. 
We investigate this further by undertaking a brute force grid search in the discretized search space of alloy recycling mixtures, see Fig.~\ref{fig_results_mo_3}. Therefore, a space grid corresponding to a 5 \% resolution of the recycling ratios is introduced. The grid search uncovers all mixtures, visualized in Fig.~\ref{fig_results_mo_3}a, including "quaternary" mixtures, where all four mixing alloys contribute with non-zero fraction, less versatile "ternary" mixtures, with only three relevant mixing alloys, and respective "binary" mixtures. The color codes in Fig.~\ref{fig_results_mo_3}a classifies the four types of different ternary recycling mixtures, while the underlying gray dots represent all recycling mixtures. Fig.~\ref{fig_results_mo_3}b visualizes the corresponding Pareto fronts. The black full line represents the Pareto front of all mixtures which beats all possible ternary mixtures in terms of strong domination in the range of TC of 169 - 182 W/(mK) and YS of 296 - 316 MPa. In this area, we provide a clear justification for facilitating the more complex recycling process that includes all four mixing alloys, compared to blending only two or three. The reason for this behaviour lies in the thermodynamics of the determined properties, which depend non-linearly on the contained 12 chemical elements. By changing the alloy mixing ratio, we vary these indirectly and in a complex way.

\section{Discussion}

Binary encoding and one-hot encoding are two of the most prominent types of 0-1 encoding \cite{Yarkoni2022}. The former is attractive due to its highly compact (i.e. qubit-sparing) coding, while the main advantage of the latter lies in the simple qubit-unique representation of the system variables which, however, comes with the price of consuming exponentially more qubits and relying on additional constraints unnecessary in binary encoding. Only one of the two encoding types (binary or one-hot) is used in each of the recently reported FM+QA studies \cite{Kitai2020,Kim2022,Kim2023,Urushihara2023,Nawa2023}.  Our studies directly compare both types applied to the method. On the one hand, it could be assumed that a less qubit consuming setup leads to better performance of the overall method due to a smaller search space. On the other hand, the compact binary encoding could potentially overload the FM approach with regard to its quadratic limitation. This would in turn counteract the beneficial qubit “economy” by demanding a much higher number of active learning iterations in comparison to the one-hot encoding case.  Overall, our results show that the drawbacks of one-hot encoding dominate in the present FM+QA workflow, whereas binary encoding provides a more efficient and robust representation for the studied alloy-optimization problems.

When using one-hot encoding, difficulties of FM+QA against FM+SA become apparent in the range of 45 to 60 logical qubits. Besides unfavorable scaling of the search space, solution search is considerably more difficult using one-hot encoding due to specific feasibility constraints added to the QUBO. This results in many local minima on the potential energy surface of the Hamiltonian. Although QA  is often expected to  be superior to SA due to its temperature-independent ability to explore the entire Hilbert space at once starting from uniform superposition state \cite{Arai2023}, our FM+QA results provide no evidence of such advantageous quantum effect. Presumably, the counteracting technical weaknesses, e.g., dissipation and decoherence, are already too dominant in the regimes for which SA performance starts to degrade. Nevertheless,  our results suggest that the QA device can act as a promising sampler for the studied workflow, and future hardware improvements may help clarify whether such samplers can provide practical advantages over classical heuristics for larger instances.

In literature, data-driven and active-learning-based QUBO quantum optimization methods like FM+QA have been only applied yet for SOO tasks, though the way to scalarize a MOO problem (i.e., to merge multiple QUBO models into a single one) is very well known \cite{Imanaka2021, Pereira2022, Schworm2024}.
The here presented non-convex MOO results from FM+QA on quantum hardware are, to the best of our knowledge,  the first of its kind, and the investigated alloy mixture system consists of 20 logical qubits, making it one of the most extensive FM+QA studies to date.

Pareto optimizers are important tools for metal recycling and inverse alloy design. In such applications, the computational cost of the optimization step often determines how many alloy constituents, process constraints, and target properties can be considered simultaneously. Expanding this design space is highly relevant in practice, as it enables the identification of recycling routes that better balance performance, resource efficiency, and sustainability. Consequently, more efficient MOO methods are of considerable interest for accelerated alloy development and circular materials design.

Our results highlight the potential of FM+QA as a data-driven Pareto black-box optimizer. This is particularly interesting because the FM+QA in Pareto search mode is generally more complicated. In fact, the QUBO model to be solved in each iteration depends not only on the training capacity but changes also with the Pareto weights. Thereby, discovering alloy mixture designs in non-convex regions crucially depends on utilizing the DDTS ansatz \cite{Plehn2025}. Restricting attention to convex regions of the Pareto front excludes many Pareto-optimal solutions. In our analysis, we discovered the most complex Pareto optimal recycling mixtures involving all available alloy components in the non-convex Pareto front region, thus discovering solutions particularly challenging to identify through human intuition or random sampling approaches.

\section{Methods}

\subsection{Dataset generation}
\label{sec_methods_data}

\begin{table}
\centering
\begin{adjustbox}{}
\begin{tabular}{ |c||c|c|c|c|c|  }
 \hline
 \multicolumn{6}{|c|}{Materials properties of mixing alloys} \\
 \hline
 Property &  \textit{Alloy 1} & \textit{Alloy 2} & \textit{Alloy 3} & \textit{Alloy 4} & \textit{Alloy 5}\\
 \hline
 Yield strength [MPa] & 436 & 273 & 338  & 253 & 420 \\
 Thermal conductivity [W/(mK)]  & 148 & 178 & 146 & 193 & 151 \\
 \hline
\end{tabular}
\end{adjustbox}
\caption{Materials properties of pure mixing alloys being our recycling ingredients. Alloy mixtures are optimized regarding yield strength maximization (SOO, Sec.~\ref{sec_res_so}) as well as maximization of thermal conductivity and yield strength (Pareto MOO, Sec.~\ref{sec_res_mo}). (All values are derived within the computational modeling elaborated in Sec.~\ref{ThermoCalc}.)}
\label{tab_2}
\end{table}

Material property data were generated using Thermo-Calc software version 2025a \cite{Andersson2002-uu} with the thermodynamic database for Al-alloys version 7, TCAL7 \cite{thermocalcThermodynamicDatabases}, and the Thermo-Calc Python interface for batch calculations \cite{scribdClientChallenge}. 
The different chemical compositions of 12 elements derived from mixtures of the five alloys introduced in Sec.~\ref{sec_res} were used as input data to calculate several alloy properties. By that, material property data for initializing the FM+QA active learning were generated. Two of the most relevant properties, YS and TC, were selected as the optimization objectives. 
YS is a fundamental property required for structural alloys, while TC is necessary for good processability, particularly in additive manufacturing.

\subsection{TC and YS models as implemented in Thermo-Calc}
\label{ThermoCalc}

Both YS and TC depend on the microstructure of alloys and, consequently, their processing history. These properties are therefore difficult to predict using machine learning methods due to the need for large datasets and physical models as well as the costly simulation times. However, Thermo-Calc offers the option of calculating material properties using physical models for individual phases and/or systems with acceptable reliability, as these have been assessed and validated for most commercial alloys.
Calculation settings are thereby configured to approximate additive manufacturing conditions, characterized by rapid solidification and particularly reduced grain sizes in the range of 0.1-1.0 $\mu$m compared to conventional as-cast states exhibiting grains sizes of up to several millimeter.
The YS can be estimated using Thermo-Calc’s property model calculator by combining phase fraction from thermodynamic calculations with physical strengthening models \cite{thermocalcAboutYield}. In principle, (total) YS, $\sigma_{y}$, is the sum of contributions from intrinsic matrix strength, $\sigma_{0}$, solid solution strengthening, $\sigma_{ss}$, grain boundary strengthening, $\sigma_{gb}$, and precipitation strengthening, $\sigma_{prec}$. The $\sigma_{0}$ describes the base strength of the matrix; in this case, the Al-FCC. The $\sigma_{ss}$ considers the introduction of elastic strains in the crystal lattice by alloying elements with a different lattice parameter, implemented by the Walbrühl model \cite{Walbruhl2017}. The $\sigma_{gb}$ is obtained using the Hall–Petch equation. Finally, $\sigma_{prec}$ describes the additional resistance caused by secondary phases. For this study, the advanced Deschamps model \cite{Deschamps1998} was used taking into account the interaction between precipitates and dislocations. 

The TC can also be calculated using the property model calculator based on the composition, phase fractions and temperature  \cite{thermocalcThermalConductivity}. In this study, the \textit{Equilibrium with Freeze-in Temperature} option was used, which freezes the thermodynamic system during cooling at a specified freeze-in temperature. This temperature was set to 400 °C. This allows non-equilibrium system states to be accounted for. The thermodynamic information obtained is then used to calculate the TC based on the Wiedemann–Franz relation \cite{Franz1853} for metallic materials, which links conductivity and electrical resistivity.

\subsection{Encoding the alloy mixture ratio}
\label{sec_methods_enc}

\begin{figure}[t]
    \centering
    \includegraphics[width=\textwidth]{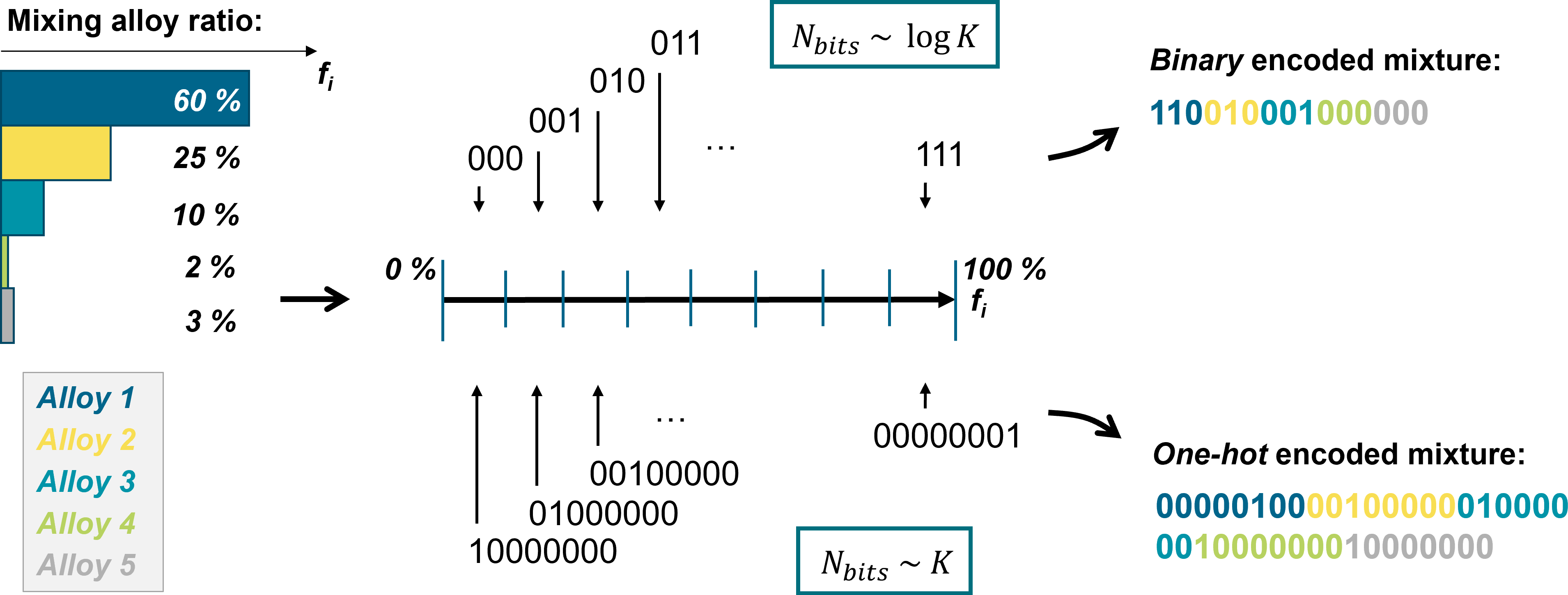}
    \caption{Illustration of the 0-1 encoding process of a recycling mixture of five mixing alloys (\textit{Alloy 1}, \textit{Alloy 2}, \textit{Alloy 3}, \textit{Alloy 4} and \textit{Alloy 5}). From left to right: Given is a specific mixture of mixing alloys in terms of five continuous alloy fractions,  $f_k \in [0,1]$. Next, in order to prepare the 0-1 encoding, each fraction range is similarly discretized into $K$ increments. Depending on the chosen type of encoding, either regular binary (upper half) or one-hot (lower half), each increment of the interval is encoded uniquely. The scaling of logical qubits consumption with the number of increments is annotated for both encoding types. The scaling advantage of the binary encoding can be realized when combining encoded fractions to the encoded alloy ratio in the last step. Please note, both types of encoding can be uniquely decoded into continuous $f_k$ again. However, due to the finite encoding fidelity, the decoded fractions have limited accuracy.}
    \label{fig_discretization}
\end{figure}

The 0-1 encoding of the mixing alloy ratios is illustrated in Fig.~\ref{fig_discretization}. Alloy ratios are defined by \textit{normalized} fractions of up to $N_\text{alloys}=5$ mixing alloys, $f_k \in [0,1]$, with $k=1,..., N_\text{alloys}$. For FM+QA, this continuous variable space must be translated into discrete binary variable space. A uniform discretization of $[0, 1]$, as described in \cite{Plehn2025}, is applied for each fraction $f_k$ determining the resolution of the mixture ratios. 
The discretized fraction axes are 0-1 encoded into bitstrings where each bit represents a logical qubit. Though several types of 0-1 encoding are known from literature, we focus on binary and one-hot encoding types \cite{Yarkoni2022, Dominguez2023}. The total number of 0-1 variables required for encoding is determined by multiplying $N_\text{qubits}$ with the number of mixing alloys $N_\text{alloys}$.

Given a certain number of qubits, $N_{\rm qubits}$, available, the encoding of a single normalized variable, $ f_k$, follows as
\begin{equation} \label{encoding}
    f_k = \sum_{i=1}^{N_{\rm qubits}} \alpha_i x_i^{(k)}
\end{equation}
Here, $x_i^{(k)} \in \{0,1\}$ is the $i$-th qubit, which encodes the fraction $ f_k$. The factors $\alpha_i$ depend on the used type of encoding. In general,  $\alpha_i$ may be $k$-dependent, however, here each $ f_k$ is processed similarly.

For the one-hot encoding, the factors result as
\begin{align}
 \alpha_i^{\text{o-h}} = i \cdot \Delta,
\end{align}
with $i=1,..., N_\text{qubits}$, where $\Delta = 1 / N_\text{qubits}$ is the increment size resulting from discretizing $ f_k$. The name one-hot encoding refers to the fact that, for this type of encoding, at most one bit can be 1 while the rest has to be 0.
In contrast to conventional one-hot encoding, $ f_k=0$ is here encoded by the zero bit-string $x_i^{(k)} = 0$ for all $i$ with the advantage of saving one bit. In this way, intuitively, each bit corresponds to a specific contribution to $ f_k$. If bit $i$ is set to 1, a part of $i/\Delta$ contributes to the total sum in Eq.~\ref{encoding}. The set of representable values spans uniformly the interval $[0,1]$.

Concerning the regular binary encoding, the factors in Eq.~\ref{encoding} read
\begin{align}
\alpha_i^\text{bin} = 2^{i-1} \cdot \Delta',
\end{align}
where again $i=1,..., N_\text{qubits}$ and the normalization constant is given by $\Delta' = 1 / (2^{N_\text{qubits}}-1)$. Hence, each bit $i$ represents a contribution of $2^{i-1} / (2^{N_\text{qubits}}-1)$ to the sum in Eq.~\ref{encoding}, i.e., the distribution of these values is not uniformly distributed across the interval $[0,1]$.

\subsection{FM+QA multi-objective active learning}
\label{sec_methods_fmqa}

In this study, the multi-objective FM+QA methodology for non-convex Pareto front discovery is employed according to previous work \cite{Plehn2025} incorporating the DDTS technique  for dataset scalarization and introducing random Pareto preferences in each active learning iteration. The multi-objective FM+QA is a follow-up of the method proposed by Kitai et al.~\cite{Kitai2020} for single-objective combinatorial optimization (see also \cite{tamura2025}).

\begin{figure}
    \centering
    \includegraphics[width=\textwidth]{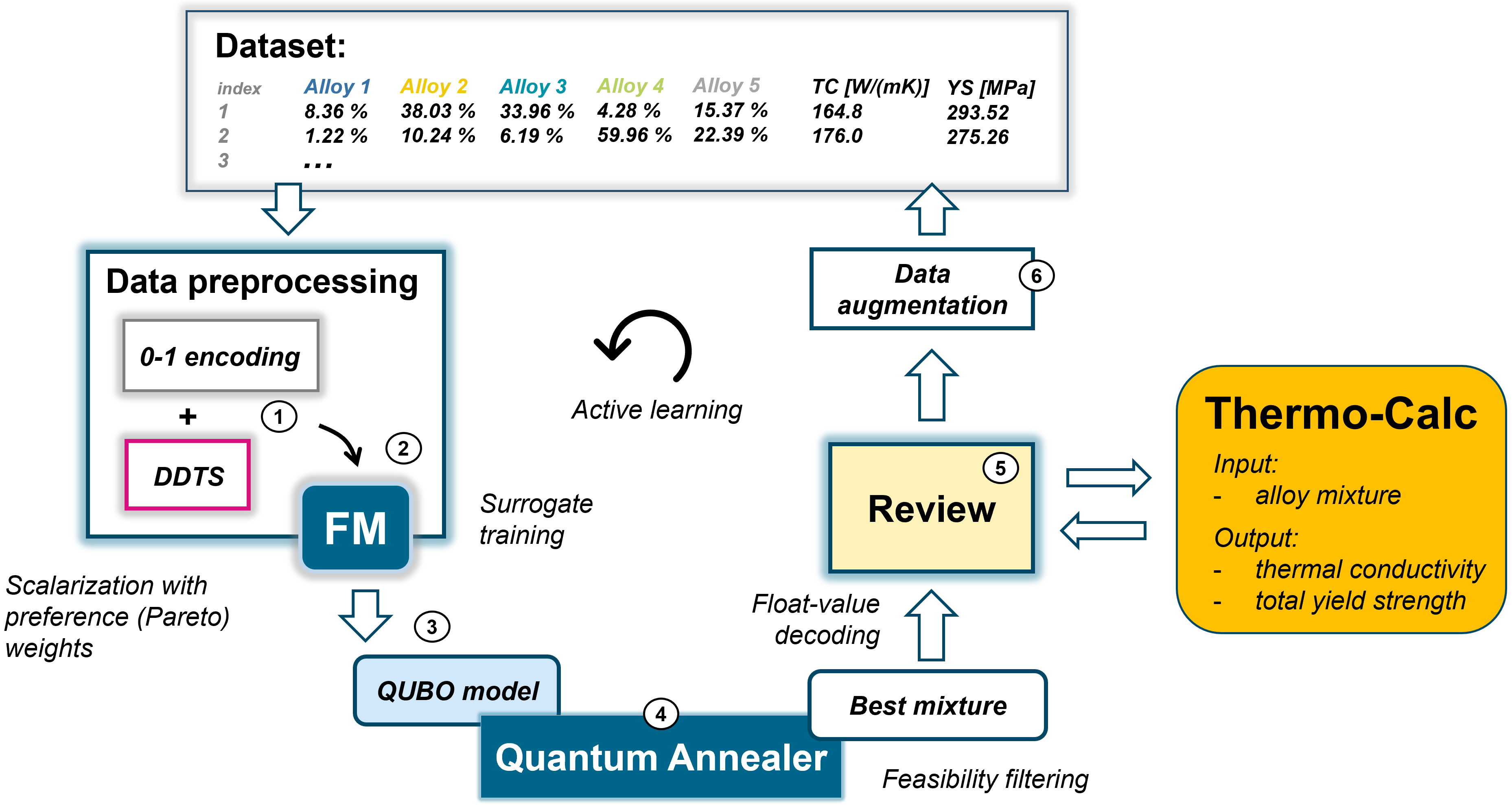}
    \caption{Multi-objective active learning FM+QA method for optimizing recycling alloy mixtures based on DDTS for scalarization. Each iteration goes through the steps: preprocessing the dataset (1) with 0-1 encoding of alloy mixtures and DDTS for scalarization; FM training (2); FM model reformulation to QUBO form (3); QUBO problem embedding on QA hardware graph and optimal recycling mixture search (4); validation of the new alloy mixture with thermodynamic simulations (5); and dataset refinement by adding the validated solution to the existing dataset (6).}
    \label{fig_fmqa_scheme}
\end{figure}

In general, the FM+QA  iterates over  data-based FM modeling, QUBO optimization and enlargement of the dataset with the validated solutions until some form of convergence is reached (see Fig.~\ref{fig_fmqa_scheme}).
Let the dataset consist of vectors, $\vec {x}$, describing the alloy mixture ratio in terms of binary variables, $x_j \in \small\{ 0,1\small\}$, and the pair of objective values $y^{(TC)}$ and $y^{(YS)}$ for material properties TC and YS, respectively. Please note, that $j=1,...N_\text{alloys} \times N_{\text{qubits}}$, because the encoded mixture ratio is obtained from concatenation of the $N_\text{alloys}$ alloy fraction encodings, $x_i^{(k)}$. The $d$-th data point follows as $(\vec{x}_d,y^{(TC)}_d, y^{(YS)}_d)$. 
In the next dataset preprocessing step, the DDTS technique is applied in order to translate the multi-objective dataset into a dataset with only a single \textit{artificial} objective. 
It follows as a simple feature mapping according to \cite{Plehn2025},
\begin{align}
(\vec{x}_d,y^{(TC)}_d, y^{(YS)}_d) \longrightarrow  (\vec{x}_d, \hat{y}_d).
\label{eq_DDTS_1}
\end{align}
where $\hat{y}_d$ is defined as
\begin{align}
\hat{y}_d = \max_{p=TC,YS} w_p (y^{(p)}_d - u_p),
\label{eq_DDTS_2}
\end{align}
attributed with different Pareto preference weights, $\vec{w}$ (with $0<w_p <1$ and $\sum_p w_p = 1$) and the so-called utopia point, $\vec u$, with $u_p < \min_d y_d^{(p)}$. 
By this procedure the dataset becomes single-objective.
Here, we refresh the DDTS within each active learning iteration by updating the utopia point and randomizing the Pareto weights for homogeneous Pareto front discovery.
After preprocessing, the FM is trained on the scalarized dataset $(\vec{x}_d, \hat{y}_d)$.

After FM training, the final QUBO problem, $f(\vec{x})$, reads
\begin{align}
f(\vec{x}) = \vec{x}^T \hat{Q} \vec{x} +\lambda C(\vec{x}) ,
\label{eq_QUBO_single_obj}
\end{align}
with the learned matrix, $\hat{Q}$, and the additionally required soft constraint term, $C(\vec{x})$. The latter is formulated quadratic in $\vec{x}$ and ensures that valid mixture ratios, according to $\sum_k f_k = 1.0$, are given preference, where the prefactor $\lambda$ tunes its strength. Please note, that another constraint term must be added in order to account for the one-hot encoding related feasibility of $\vec{x}$.
Solving $f(\vec{x})$ with QA provides an optimal alloy mixture corresponding to the present accuracy of the learned FM and describing the next most promising candidate mixture. Finally, the FM is validated and improved. Practically, it is realised by determining the "true" TC and YS values for the next candidate mixture and by adding this new, validated data point to the dataset. This review process is facilitated by thermodynamic calculations, where we utilizes the Thermo-Calc software (see Sec.~\ref{ThermoCalc}). By refining the dataset in this manner, the FM model improves iteratively. 

Please note finally that a QUBO optimizer solution which has been already found previously is substituted by a new feasible solution. In this process, exploration and exploitation is balanced by means of a global random pick and a local search mechanism. 

\subsubsection{FM training settings}
\label{sec_methods_fm}

The FM+QA employs a FM \cite{Rendle2010}, which models interactions among binary variables $x_i$ up to second order,
\begin{align} 
f^{\text{FM}}(\vec{x}) = w_0 + \sum_{i=1}^{N} w_i x_i + \sum_{i<j}^{N} w_{ij} x_i x_j, \label{eq_FM} 
\end{align} 
where $w_0$ is the bias, $w_i$ are linear weights, and $w_{ij}$ are pairwise interaction weights. The $N^2$ parameters $w_{ij}$ are determined via factorization $w_{ij} = \langle  \mathbf{v}_i, \mathbf{v}_j \rangle$, where $\mathbf{v}_i\in \mathbf{R}^{k}$
 are latent embeddings. This reduces parameters from $N^2$ to $N\cdot k$, with $k$ being the factorization rank. The interaction strength reflects embedding proximity in latent space. The matrix notation results by defining $Q = V^TV$, where $V \in \mathbf{R}^{k\times N}$, yielding $Q_{ii} = w_i$ and $Q_{ij}=w_{ij}$ for $i<j$. In this work, the fastFM \cite{Bayer2016} package is employed with alternating least squares for training, splitting data into training (80 \%), validation (16 \%) and test (4 \%) sets. Due to sensitivity of early active learning iterations to hyperparameters, the FM regularization parameters are optimized based on the validation set using Optuna \cite{Optuna2019} for 50 repetitions, while the factorization rank is fixed to 6 and training runs through maximum 1000 epochs.

\subsubsection{Optimizer settings}
\label{sec_methods_qa}

In each iteration, the FM+QA algorithm executes a QA hardware experiment on a D-Wave Advantage system \cite{Munoz2025, dwave_ocean_docs, dwave_advantage} to find the ground state of the Ising model corresponding to the trained QUBO problem after embedding onto the QA hardware graph. 
Several parameters influence QA performance and solution quality. Here, annealing time, intra-chain coupling strength and constraint penalty strengths (applied to the QUBO after FM training) are systematically configured in order to enhance QA hardware performance. The number of anneals and spin-reversal transformations for error correction are specified for each sub study as detailed in Sec.~\ref{sec_res}. For the same purpose, the assignment of logical qubits in the hardware embedding is shuffled several times within each QA experiment, which is straightforwardly implemented due to QUBO full connectivity.

In general,  automated active learning motivates a QA hardware configuration that is sufficient across all iterations avoiding the computational overhead of an iteration-wise configuration. In this work, the QA calibration is performed specifically for each sub study, where three typical QUBO problem instances are randomly selected from previous FM+SA runs, in particular, from early, mid, and late stages of the active learning process (e.g., iterations 19, 64 and 134). Thereby, it is assumed that QUBOs from corresponding FM+SA runs approximately represent typical QUBO problems the QA has to tackle during the FM+QA run. In fact, in corresponding FM+QA and FM+SA runs, the alloy mixture system, the dataset, and the FM modeling are similar. Notably, however, this approach requires prior FM+SA analogues.
The calibration is then based on conducting two test series for each selected QUBO instance: First, various penalty strengths are scanned while fixing the annealing time. The best value is identified by evaluating the average probability of returning feasible solutions and the total exact solution count. Second, various annealing time values are tested applying the previously determined optimal penalty strength and using the same evaluation criteria. For the small-scale studies, penalty strength calibration are finally repeated based on the optimal annealing time to verify consistency.
In particular, annealing times of 1.0, 2.5, 5.0, 10, 20 and 40 µs were tested, with the setting of 2.5 µs proving to be the most economical in all cases.

The FM+SA algorithm solves the QUBO problem by means of SA. In this work, the SA algorithm is implemented based on the D-Wave Ocean software suite \cite{dwave_ocean_docs}. For all our FM+SA studies, the number of annealing runs and sweeps per SA optimization are kept fixed at 1000 and 2000, respectively. However, the SA schedule is automatically adjusted to each specific QUBO problem. For this purpose, the Ocean internal \textit{\_default\_ising\_beta\_range()} sampler routine is utilized, which determines the temperature schedule based on the maximum and minimum effective bias per Ising spin variable. This ensures rapid equilibration at the start (high temperature) and minimal excitations at the end (low temperature).

\section{Conclusions}
In this study, we assessed factorization machine-based data-driven multi-objective optimization (MOO) for alloy recycling using quantum annealing (QA). We carried out a direct comparison with its classical simulated annealing (SA) competitor under matched active learning conditions. The results show that QA can be applied not only to single-objective optimization (SOO), but also to MOO alloy design tasks when combined with the data-driven Tchebycheff scalarization (DDTS) technique. In the investigated alloy recycling systems, DDTS enabled the discovery of Pareto-optimal recycling mixtures along both convex and non-convex regions of the Pareto front, including mixtures that involve all considered alloy components. 

For yield strength SOO, binary encoding consistently outperformed one-hot encoding. This advantage is most likely due to the much smaller number of required logical qubits and therefore the more efficient sampling and surrogate (FM) model training. With binary encoding, FM+QA and FM+SA showed comparable optimization performance up to the largest investigated system with 25 logical qubits. 

For the MOO case, we demonstrated FM+QA combined with DDTS for Pareto optimization of alloy mixtures. Specifically, when Pareto-maximizing yield strength and thermal conductivity, the method produced a well-resolved Pareto front, including non-convex regions, which are difficult to access with classical weighted-sum scalarization. The comparison with FM+SA showed only minor differences in Pareto front quality and hypervolume progression, indicating that the D-Wave Advantage QA device can serve as a competitive sampler.

The time-to-solution (TTS) analysis further indicates that, for the studied problem sizes up to 25 logical variables with binary encoding,  QA delivered competitive, and in some cases lower, TTS than SA under the workflow-specific settings considered here. However, this observation should not be interpreted as a general quantum speedup, since neither solver was exhaustively tuned for asymptotic benchmarking.

The main limitations observed for the QA approach are the sampling quality and the calibration effort required to tune the quantum hardware. In particular, one-hot encoded problems become challenging between 45 and 60 logical qubits, where embedding overhead, feasibility constraints, and hardware noise strongly affect FM+QA performance. The results therefore suggest that, at least for the SOO cases, the present bottleneck is not the FM+QA formulation itself, but the practical difficulty of obtaining reliable samples from calibrated quantum hardware, in particular, concerning automated high-throughput approaches. 
In this regard, implementing approximate optimization schemes on gate-based quantum hardware architectures, such as the QAOA algorithm, might be an alternative promising perspective, which provides more versatile calibrations along the annealing schedule and practices for inherently obeying one-hot constraints.

A further limitation of the present workflow lies in the use of Thermo-Calc as the feedback engine within the active learning loop. Although Thermo-Calc provides a powerful and industry-relevant framework for estimating alloy properties based on thermodynamic and physically informed models, its predictions remain model-dependent and rely on the validity of the underlying databases. In particular, quantities such as yield strength and thermal conductivity are sensitive to processing history, microstructure evolution, and non-equilibrium effects that may not be fully captured in the present setup. First and foremost, each feedback evaluation has very high computational costs, which limits the total number of active learning iterations. This highlights the importance of efficient active learning strategies that maximize information gain from as few evaluations as possible. Such efficiency becomes even more critical when replacing computational feedback with manufacturing and experimental validation pipelines, where each iteration may require substantial time, material resources, and labor. Future work should therefore incorporate experimental validation, higher-fidelity simulations, or uncertainty-aware feedback models to further strengthen the optimization loop.

Overall, the study supports a cautious but positive outlook for FM+QA as a data-driven MOO tool for materials design and specifically alloy up-cycling. Future progress will likely depend on improved QA hardware connectivity, reduced thermal noise, more robust embedding and automated calibration strategies as well as strategies on constraint handling. A promising future direction could be the development of hybrid strategies that combine QA as an efficient sampler with classical local search, refinement, and constraint-handling routines, thereby exploiting the complementary strengths of both approaches.

\section{Data and Code Availability}

The source code of the single and multi-objective FM+QA algorithm is employed according to reference \cite{Plehn2025} and available on GitHub: \url{https://github.com/dlr-wf/fmqo-alloy-discovery}.
The data supporting this study were generated using the commercial software Thermo-Calc. Due to licensing restrictions associated with this software, the raw data cannot be made publicly available. Processed data and additional information may be provided by the authors upon reasonable request.

\section{Acknowledgement}
We acknowledge the financial support of the DLR Quantum Computing Initiative through the project QuantiCoM (\url{https://qci.dlr.de/quanticom/}) funded by the German Federal Ministry of Research, Technology and Space (BMFTR).
The authors gratefully acknowledge the Jülich Supercomputing Centre (\url{https://www.fz-juelich.de/jsc}) for funding this project by providing computing time on the D-Wave Advantage™ System JUPSI through the Jülich UNified Infrastructure for Quantum computing (JUNIQ).

\section{Author Contributions}
T.~P.~led the project, implemented the FM+QA workflow, performed the data analysis. K.~B.~developed the Thermo-Calc and TC-Python workflow. K.~B.~and S.~T.~guided the alloy recycling use case. D.~B.-Y.~and D.~M.~contributed to the conceptual development and supervised the research. All authors wrote and reviewed the manuscript.

\section{Competing Interests}
T.~P., D.~M., D.~B.-Y.~are named inventors in the pending patent application DE102025147173.8, filed by the German Aerospace Center (DLR), related to the results described in this work.

\bibliographystyle{naturemag}
\bibliography{bibliography}

\end{document}